\documentclass[%
reprint,
 amsmath,amssymb,
 aps,
prc,
]{revtex4-2}

\usepackage{graphicx}% Include figure files
\usepackage{dcolumn}% Align table columns on decimal point
\usepackage{bm}% bold math
\usepackage{hyperref}% add hypertext capabilities
\usepackage{physics}
\usepackage{epstopdf}
\usepackage{epsfig}
\usepackage{longtable}
\usepackage[mathlines]{lineno}% Enable numbering of text and display math
\begin{document}

\preprint{APS/123-QED}

\title{Exchange correction for allowed $\beta$-decay}% Force line breaks with \\

\author{O. Ni\c{t}escu$^{1,2,3}$}

%\author{M.I. Krivoruchenko$^{4,5}$}

\author{S. Stoica$^{3}$}

\author{F. \v{S}imkovic$^{1,4,5}$}
\email{fedor.simkovic@fmph.uniba.sk}

\affiliation{$^{1}$Faculty of Mathematics, Physics and Informatics, Comenius University in Bratislava, 842 48 Bratislava, Slovakia}

\affiliation{$^{2}$“Horia Hulubei” National Institute of Physics and Nuclear Engineering, 30 Reactorului, POB MG-6, RO-077125 Bucharest-M\u{a}gurele, Romania}
\affiliation{$^{3}$ International Centre for Advanced Training and Research in Physics, P.O. Box MG12, 077125 M\u{a}gurele, Romania}

%\affiliation{$^{4}$Institute for Theoretical and Experimental Physics, NRC "Kurchatov Institute",
%	B. Cheremushkinskaya 25, 117218 Moscow, Russia}
%\affiliation{$^{5}$National Research Center "Kurchatov Institute", Ploshchad' Akademika Kurchatova 1, 123182 Moscow, Russia}

\affiliation{ $^{4}$Bogoliubov Laboratory of Theoretical Physics, Joint Institute for Nuclear Research, 141980 Dubna, Russia}

\affiliation{ $^{5}$Institute of Experimental and Applied Physics, Czech Technical University in Prague,
	110 00 Prague, Czech Republic}

\iffalse

\collaboration{MUSO Collaboration}%\noaffiliation

\author{Charlie Author}
 \homepage{http://www.Second.institution.edu/~Charlie.Author}
\affiliation{
 Second institution and/or address\\
 This line break forced% with \\
}%
\affiliation{
 Third institution, the second for Charlie Author
}%
\author{Delta Author}
\affiliation{%
 Authors' institution and/or address\\
 This line break forced with \textbackslash\textbackslash
}%

\collaboration{CLEO Collaboration}%\noaffiliation

\fi

\date{\today}% It is always \today, today,

\begin{abstract}
We investigate the exchange effect between the final atom's bound electrons and those emitted in the allowed $\beta$-decay of the initial nucleus. The electron wave functions are obtained with the Dirac-Hartree-Fock-Slater self-consistent method, and we ensure the orthogonality between the continuum and bound electron states, in the potential of the final atom, by modifying the last iteration of the self-consistent method. We show that orthogonality plays an essential role in calculating the exchange correction. After imposing the orthogonality, we found considerable differences in magnitude and energy dependence compared to previous results. We argue that our findings can solve the mismatch between the previous predictions and experimental measurements in the low-energy region of the $\beta$ spectrum. First, we calculate the exchange effect for the low-energy $\beta$ transitions in $^{14}$C, $^{45}$Ca, $^{63}$Ni, and $^{241}$Pu, recently investigated in the literature. Next, we compute the total exchange correction for a large number of $\beta$ emitters, with $Z$ from $1$ to $102$. From the systematic study, we found that for ultra-low energy, i.e., $5$ eV, the $Z$ dependence of total exchange effect is affected by $s_{1/2}$ and $p_{1/2}$ orbitals closure. We also show that the contributions from orbitals higher than $2s_{1/2}$ orbital are essential for correctly calculating the total effect, especially for low energies and heavy $\beta$ emitters. Finally, we provide an analytical expression of the total exchange correction for each atomic number for easy implementation in experimental investigations. 

\iffalse
\begin{description}
\item[Usage]
Secondary publications and information retrieval purposes.
\item[Structure]
You may use the \texttt{description} environment to structure your abstract;
use the optional argument of the \verb+\item+ command to give the category of each item. 
\end{description}
\fi

\end{abstract}

\maketitle

%\tableofcontents

\section{Introduction}

The continuous development of the experimental techniques for measuring the nuclear $\beta$ decay spectra has recently increased the interest in calculating the atomic corrections for the process. Precise investigations of the $\beta$ spectra are currently used for getting information on some fundamental, still unknown, issues. For example, one can answer the most pressing question in particle physics: What is the absolute value of the neutrino mass? The most sensitive experiments aiming to measure the neutrino mass are based on $\beta$ transitions with low kinetic energy released in the decay. The Karlsruhe tritium neutrino mass (KATRIN) experiment, using the superallowed $\beta$ decay of tritium, recently reported the best upper bound, $m_\beta \leqslant 0.8$ eV \cite{KATRIN-N2022}. Precise measurements of the $\beta$ spectra shape could also uncover the existence of exotic currents beyond the standard electroweak model \cite{HerczegPPNP2001,NicoARNPS2005,Severijns-RMP2006,Severijns-ARNPS2011,Severijns-JPG2014,Hayen-RMP2018}. Lorentz invariance violation can also be investigated by searching for sideral variations in the $\beta$ decay spectra \cite{Noordmans-PRC2015}. Not only that the study of $\beta$-decay has solved fundamental physics problems, but it can still solve some of them, mainly in neutrino physic. Therefore, predictions of the $\beta$ spectra should include all relevant corrections to match the experimental precision \cite{Hayen-RMP2018}.

The primary atomic effects on the $\beta$ spectrum arise from the screened $\beta$ particle wave function due to the atomic electron cloud, the exchange between the emitted and bound electrons, and the sudden change of the nuclear charge. The latter can induce internal ionization (shake-off) or atomic excitations (shake-up), and its effect is not expected to exceed $0.1\%$. Contrary, the screening, and the exchange corrections can significantly influence the $\beta$ spectrum shape and the decay rate, especially for low energetical transitions. The exchange correction arises from creating a $\beta$ electron in a bound orbital of the final atom corresponding to one occupied in the initial atom. Simultaneously, an atomic electron from the bound orbital makes a transition to a continuum orbital of the final atom.     

The study of the exchange correction started in 1962 in Bahcall's paper \cite{Bahcall-PR1963}, which was only focused on the interchange of the emitted electrons from different $\beta$ transitions with the bound electrons from only the $1s$ orbital. The approximation led to a lower emission probability at low energy. Twenty years later, Haxton showed that the exchange correction for tritium $\beta$ decay enhances the process at low energy \cite{Haxton-PRL1985}. The latter conclusion is in agreement with the recent experimental measurements. The enhancement at low energy was also confirmed by Harston and Pyper \cite{Harston-PRA1992} in the case of multiple nuclear $\beta$ decays, the allowed $\beta$ transitions of $^{14}$C, $^{35}$S, $^{106}$Ru and the nonunique first forbidden $\beta$ transition of $^{241}$Pu.

Recently, the exchange correction has been revisited for the nonunique first forbidden $\beta$ decay of $^{241}$Pu \cite{Mougeot-PRA2012}. Although the wave functions for the emitted and bound electrons were the analytical solutions of the Dirac equation in the hydrogenic approximation given by Rose \cite{Rose-1961}, it was clear that the electron spectrum, with the exchange correction, better fits the experimental values. The results for $^{241}$Pu were refined with more sophisticated wave functions in \cite{Mougeot-PRA2014}, where the allowed $\beta$ transition of $^{63}$Ni is also presented. In a review of the analytical corrections for allowed $\beta$ decays \cite{Hayen-RMP2018}, it can be found some further investigations on the exchange corrections for the decay of $^{45}$Ca and $^{241}$Pu. Besides the matching between the predicted and the experimental spectrum, the exchange effect is also important when excluding background $\beta$ events from sensitive experimental analysis, e.g., LUX-ZEPLIN \cite{LUX-ZEPLINPRD2020}, XENONnT \cite{AprileJCAP2020}, XENON1T \cite{XENON-PRD2020}. In this regard, the atomic exchange correction was investigated for the unique first forbidden transition of $^{85}$Kr and the nonunique first forbidden transitions of $^{212}$Pb and $^{214}$Pb \cite{Haselschwardt-PRC2020}.   

Considering the increasing interest in the field and current experimental capabilities of measuring the low-energy region of $\beta$ spectra, for example, with metallic magnetic calorimeters (MMCs) \cite{LoidlARI2014,LoidlARI2019,KossertARI2022}, we reexamine the exchange effect calculation. For the electron wave functions, we employ a modified Dirac-Hartree-Fock-Slater self-consistent method, which ensures the orthogonality between the continuum and bound electron states in the potential of the final atom. We found that the final atom's non-orthogonal continuum and bound states lead to undesired errors in the overlaps between the initial atom's bound states and the final atom's continuum states, which translates into a downturn in the total exchange correction. We present the calculations in detail for the $\beta$ decays of $^{14}$C, $^{45}$Ca, $^{63}$Ni and $^{241}$Pu. Then we perform the exchange correction calculation for a large number of $\beta$ emitters, with an atomic number in the range $Z=1-102$. Except for the low-energy region, the total exchange correction progressively increases with the nuclear charge. Still, even for $Z=100$, the total exchange effect is under $1\%$ at $200$ keV kinetic energy of the emitted electron so that the correction can be completely neglected after this region, depending on the aimed accuracy. We also investigate how the partial contributions, coming from orbitals higher than $2s_{1/2}$ orbital, influence the calculation of the total effect. Those partial contributions are important, especially for heavier atoms, and their impact decreases with the increasing energy of the emitted electron. Finally, we provide an analytical expression of the total exchange correction for each atomic number for easy implementation in experimental investigations.

\section{Atomic Exchange Correction for allowed $\beta$ transitions}

The electron spectrum for a nuclear $\beta$ decay is proportional with \cite{BehrensBook1982}

\begin{equation}
\label{eq:electronSpectrum}
\frac{d\Gamma}{dE_e}\propto p_e E_e (E_0-E_e)^2 F_0(Z',E_e)C(E_e),
\end{equation} 
where the term $p_e E_e (E_0-E_e)^2$ is related to the statistical phase space factor that reflects the momentum distribution between the neutrino and the electron, $F_0(Z',E_e)$ is the so-called Fermi function and the shape factor, $C(E_e)$, contains the nuclear matrix elements and all the possible remaining terms dependent on lepton energy. Here $E_e$ is the total energy of the electron, $p_e=\sqrt{E_e^2-m_e^2}$ is the electron momentum, $E_0=Q+m_e$ is the maximum energy of the electron and $Z'$ is the atomic number of the final nucleus.

The Fermi function, which encodes the electrostatic interaction between the electron and the final atom, can be expressed in terms of the large- and small-component radial functions, $g_\kappa(E_e,r)$ and $f_\kappa(E_e,r)$ of the emitted electron,

\begin{equation}
\label{eq:FermiFunction}
F_0(Z',E_e) = {g}^2_{-1}(E_e,R) + {f}^2_{+1}(E_e,R),
\end{equation}    
evaluated on the nuclear surface of the final nucleus. The radial components of the continuum states and their normalization are discussed in the following section.

A widely used approximation for the Fermi function is obtained by keeping the lowest power of the expansion in $r$ of the radial wave functions corresponding to a uniformly charged sphere potential \cite{DoiPTPS1985},  

\begin{eqnarray}
F_{k-1}(Z',E_e)&=&\left[\frac{\Gamma(2k+1)}{\Gamma(k)\Gamma(2\gamma_k+1)}\right]^2(2pR)^{2(\gamma_k-k)}e^{\pi\eta} \nonumber\\ 
&&\times \left|\Gamma(\gamma_k+i\eta)\right|^2,
\end{eqnarray}
with $k=1$ for allowed $\beta$ transitions.
The remaining quantities are given by
\begin{align}
\begin{aligned}
\gamma_k&=\sqrt{k^2-(\alpha Z')^2},\\
\eta&=\alpha Z' \frac{E_e}{p_e},
\end{aligned}
\end{align} 
and $\Gamma(z)$ is the Gamma function.

For allowed transitions, the shape factor is not energy dependent, so their spectra are proportional with
\begin{equation}
\frac{d\Gamma}{dE_e}\propto p_e E_e (E_0-E_e)^2 F_0(Z',E_e).
\end{equation}

The modification of the allowed spectrum due to the exchange correction is given by the following transformation \cite{Harston-PRA1992} 
\begin{equation}
\frac{d {\Gamma}}{dE_e}\Rightarrow\frac{d {\Gamma}}{dE_e}\times\left[1+\eta^T(E_e)\right]
\end{equation}
where 
\begin{eqnarray}
\label{eq:TotalEchangeCorrection}
\eta^T(E_e)&=&f_s(2T_{s}+T_{s}^2)+(1-f_s)(2T_{\bar{p}}+T_{\bar{p}}^2)\nonumber\\
&=&\eta_{s}(E_e)+\eta_{\bar{p}}(E_e)
\end{eqnarray}
Here,
\begin{eqnarray}
f_s=\frac{g'^2_{-1}(E_e,R)}{g'^2_{-1}(E_e,R)+f'^2_{+1}(E_e,R)},
\end{eqnarray}
and the quantities $T_{s}$ and $T_{\bar{p}}$ depend respectively on the overlaps between the bound $s_{1/2}$ ($\kappa=-1$) and $\bar{p}\equiv p_{1/2}$ ($\kappa=1$) orbitals wave functions in the initial state atom and the continuum states wave functions in the final state atom,

\begin{eqnarray}
\label{eq:TnsQuantities}
T_{s}=\sum_{(ns)'}T_{ns}=-\sum_{(ns)'}\frac{\braket{\psi'_{E_es}}{\psi_{ns}}}{\braket{\psi'_{ns}}{\psi_{ns}}}\frac{g'_{n,-1}(R)}{g'_{-1}(E_e,R)}
\end{eqnarray}
and
\begin{eqnarray}
T_{\bar{p}}=\sum_{(n\bar{p})'}T_{n\bar{p}}=-\sum_{(n\bar{p})'}\frac{\braket{\psi'_{E_e\bar{p}}}{\psi_{n\bar{p}}}}{\braket{\psi'_{n\bar{p}}}{\psi_{n\bar{p}}}}\frac{f'_{n,+1}(R)}{f'_{+1}(E_e,R)}.
\end{eqnarray}
The sums are running over all occupied orbitals of the final atom, which, in the sudden approximation, correspond to the parent electronic configuration. The radial components of the bound states, $g_{n,\kappa}(r)$ and $f_{n,\kappa}(r)$, are discussed in the following section. All primed wave functions refer to the final atom.

According to the definition of the quantities $T_{s}$ and $T_{\bar{p}}$, we can write
\begin{equation}
\label{eq:PartialExchnageSOrbitals}
\eta_{s}(E_e)=\sum_{n}\eta_{ns}+f_s\sum\limits_{\substack{n,m \\ n\neq m}}T_{ns}T_{ms}
\end{equation}  
and 
\begin{equation}
\eta_{\bar{p}}(E_e)=\sum_{n}\eta_{n\bar{p}}+(1-f_s)\sum\limits_{\substack{n,m \\ n\neq m}}T_{n\bar{p}}T_{m\bar{p}}
\end{equation}
Here we defined the partial exchange correction of $n$th $s$ orbital and $n$th $\bar{p}$ orbital as,
\begin{equation}
\eta_{ns}=f_s(2T_{ns}+T^2_{ns})
\end{equation}
and
\begin{equation}
\eta_{n\bar{p}}=(1-f_s)(2T_{n\bar{p}}+T^2_{n\bar{p}}),
\end{equation}
respectively.

Although the contribution coming from the exchange with $p_{1/2}$ orbitals, $\eta_{\bar{p}}(E_e)$, is small, it should not be neglected if we aim for high precision \cite{Hayen-RMP2018}. Depending on the specific application, the $p_{1/2}$ contribution was ignored \cite{Harston-PRA1992,Mougeot-PRA2012,Mougeot-PRA2014} or included \cite{Hayen-RMP2018,Haselschwardt-PRC2020}. Moreover, the mixed sum in Eq.~\ref{eq:PartialExchnageSOrbitals} was not taken into account in \cite{Harston-PRA1992} for the reason that its contribution is small in $\eta_{s}(E_e)$. The approximation is motivated for high energies, but its contribution can be about $1-3\%$ for kinetic energies below $1$ keV \cite{Mougeot-PRA2012}.

The essential quantities in computing the exchange correction are the overlaps between the bound orbitals electron wave functions in the initial atom and the continuum state electron wave function, with energy $E_e$, in the final atom, $\braket{\psi'_{E_es}}{\psi_{ns}}$. In numerical calculation, we must overcome three difficulties. The first two are related to the oscillatory nature of the continuum state wave function. First, good knowledge of the wave function is required over a wide region of space, and second, special care should be addressed for the integration method. The last and the most important difficulty is to ensure that the final-state electron continuum wave function is orthogonal to the wave functions of the final-state bound orbitals, e.g., $\braket{\psi'_{E_es}}{\psi'_{ns}}=0$. The overlap should be zero because its wave functions are eigenfunctions of the same Hamiltonian \cite{Harston-PRA1992}. If the overlap integral $\braket{\psi'_{E_es}}{\psi'_{ns}}$ is not zero or at least much smaller then $\braket{\psi'_{E_es}}{\psi_{ns}}$, then the overlap $\braket{\psi'_{E_es}}{\psi_{ns}}$ may be significantly in error \cite{Harston-PRA1992}. The effect was also studied in different nucleon removal reactions, $(\gamma,p)$ and $(e,e'p)$, and significant modifications were found in the polarization of the outgoing protons and photon asymmetry for $(\gamma,p)$ reaction \cite{JohanssonNPA2000}.

 To verify the orthogonality, we define as reference the following dimensionless quantities,

\begin{eqnarray}
\label{eq:TrefnsQuantities}
T^{\textrm{ref}}_{ns}=-\frac{\braket{\psi'_{E_es}}{\psi'_{ns}}}{\braket{\psi'_{ns}}{\psi_{ns}}}\frac{g'_{n,-1}(R)}{g'_{-1}(E_e,R)},
\end{eqnarray}
and
\begin{eqnarray}
T^{\textrm{ref}}_{n\bar{p}}=-\frac{\braket{\psi'_{E_e\bar{p}}}{\psi'_{n\bar{p}}}}{\braket{\psi'_{n\bar{p}}}{\psi_{n\bar{p}}}}\frac{f'_{n,+1}(R)}{f'_{+1}(E_e,R)},
\end{eqnarray}
which should be zero for any energy of the emitted electron.

\section{Electron Wave functions}

The electron relativistic wave function for a spherically symmetric potential, $V(r)$, is defined as \cite{RoseBook1961}
\begin{eqnarray}
\label{eq:generalSolution}
\psi(E_e,\boldsymbol{r})=\sum_{\kappa m}^{}\begin{pmatrix}
g_\kappa(E_e,r)\Omega_{\kappa, m}(\hat{\boldsymbol{r}})\\
i f_\kappa(E_e,r)\Omega_{-\kappa, m}(\hat{\boldsymbol{r}})
\end{pmatrix},
\end{eqnarray}
where $g_\kappa$ and $f_\kappa$ are the large- and small-component radial functions, respectively, which satisfy the following system of coupled differential equations
\begin{eqnarray}
\label{eq:radialEquations}
\left(\frac{d}{dr}+\frac{\kappa+1}{r}\right)g_\kappa-(E_e-V(r)+m_e)f_\kappa&&=0,\nonumber\\
\left(\frac{d}{dr}-\frac{\kappa-1}{r}\right)f_\kappa+(E_e-V(r)-m_e)g_\kappa&&=0,
\end{eqnarray}
where $E_e$ is the total electron energy. Here, $\boldsymbol{r}$ stands for the position vector of the electron, $r=\left|\boldsymbol{r}\right|$ and $\hat{\boldsymbol{r}}=\boldsymbol{r}/r$. The relativistic angular momentum quantum number, $\kappa$, takes positive and negative integer values, and specifies both the total angular momentum, $j$, and the orbital angular momentum, $\ell$, by
\begin{eqnarray}
j=\left|\kappa\right|-1/2, \hspace{0.2cm} \ell=\begin{cases}
\kappa  &\quad \text{if } \kappa>0,\\
\left|\kappa\right|-1  &\quad \text{if } \kappa<0.\\
\end{cases}
\end{eqnarray}
The spherical spinors, $\Omega_{\kappa, m}(\hat{\boldsymbol{r}})$, are defined by \cite{RoseBook1995,VarshalovichBook1988}
\begin{eqnarray}
\Omega_{\kappa, m}(\hat{\boldsymbol{r}})=\sum_{\mu=\pm1/2}\braket{\ell,\frac{1}{2},m-\mu,\mu}{j,m}Y_{\ell,m-\mu}(\hat{\boldsymbol{r}})\chi_\mu\nonumber\\
\end{eqnarray} 
where $\braket{j_1,j_2,m_1,m_2}{j,m}$ are Clebsch-Gordan coefficient, $Y_{\ell,m}(\hat{\boldsymbol{r}})$ are the spherical harmonics, and $\chi_\mu$ are the usual Pauli spinors.

For atomic bound states ($E_e<m_e$), each discrete energy level is characterized by its quantum number $\kappa$, its principle quantum number $n$, its binding energy $\epsilon_e^{n\kappa}$, and its total energy $E_e^{n\kappa}=m_e-\left|\epsilon_e^{n\kappa}\right|$. The bound orbitals satisfy the orthonormality relation 
\begin{eqnarray}
\braket{\psi_{n\kappa m}}{\psi_{n'\kappa' m'}}=\delta_{nn'}\delta_{\kappa\kappa'}\delta_{mm'},
\end{eqnarray} 
where $\braket{\boldsymbol{r}}{\psi_{n\kappa m}}=\psi_{n\kappa m}(\boldsymbol{r})$.
In terms of the large- and small-component radial wave functions, the scalar product can be written in the following explicit form:
\begin{eqnarray}
\braket{\psi_{n\kappa m}}{\psi_{n'\kappa' m}}=&&\int_{0}^{\infty}r^2\left[g_{n,\kappa}(r)g_{n',\kappa'}(r)\right]dr\nonumber\\
&&+\int_{0}^{\infty}r^2\left[f_{n,\kappa}(r)f_{n',\kappa'}(r)\right]dr.
\end{eqnarray}

For the calculation of the bound orbitals, we employed the \textsc{RADIAL} subroutine package \cite{SalvatCPC2019}. A comprehensive manual of the package can be found in the Supplementary Material of \cite{SalvatCPC2019}. The program DHFS.f, included in the package, solves the Dirac-Hartree-Fock-Slater (DHFS) equations for the ground-state configuration of neutral atoms and positive ions with $N_e$ bound electrons and $Z_p$ protons in the nucleus. In this section, $Z_p$ can be either the atomic number for the final nucleus, $Z'$, or the atomic number for the initial nucleus, $Z$. Although the DHFS equations are obtained by replacing the non-local exchange potential with a local exchange approximation \cite{SlaterPR1951} the results are reliable, and the procedure is efficient. There is evidence \cite{RosenPR1968} that the local exchange approximation can lead to accurate electron binding energies without the need for the extensive numerical calculations of the non-local exchange potential entailed by the full Hartree-Fock approach.

The system of differential equations Eq.~(\ref{eq:radialEquations}) is solved in the DHFS potential,
\begin{eqnarray}
V_{\text{DHFS}}(r)=V_{\text{nuc}}(r)+V_{\text{el}}(r)+V_{\text{ex}}(r)
\end{eqnarray}
which is a sum of the nuclear, electronic and exchange potentials.

 For the nuclear potential, $V_{\text{nuc}}(r)$, it is considered the electrostatic interaction of an electron at distance $r$ with a spherical nucleus filled with protons following a Fermi distribution  \cite{HahnPR1956}    
\begin{eqnarray}
\rho_p(r)=\frac{\rho_0}{1+e^{\left(r-R_n\right)/z}},
\end{eqnarray}
where $R_n=1.07A^{1/3}$ fm, $z=0.546$ fm, and $\rho_0$ must be determined from normalization. Thus, the nuclear potential is
\begin{eqnarray}
	V_\text{nuc}(r)=-\alpha\int\frac{\rho_p(r')}{|\boldsymbol{r}-\boldsymbol{r'}|}d\boldsymbol{r'}.
\end{eqnarray}

The electronic potential describes the interaction of an electron at distance $r$ with the atomic cloud, and it is found from integrating over the volume of the electron density, $\rho(r)$, 
\begin{eqnarray}
	V_\text{el}(r)=\alpha\int\frac{\rho(r')}{|\boldsymbol{r}-\boldsymbol{r'}|}d\boldsymbol{r'}.
\end{eqnarray} 

Due to Slater's approximation \cite{SlaterPR1951}, the exchange potential can be expressed in terms of the electron density in the following way
\begin{eqnarray}
V_{\text{ex}}^{\text{Slater}}(r)=-\frac{3}{2}\alpha	\left(\frac{3}{\pi}\right)^{1/3}\left[\rho(r)\right]^{1/3}.
\end{eqnarray}

\begin{figure}[h]
	\includegraphics[width=0.48\textwidth]{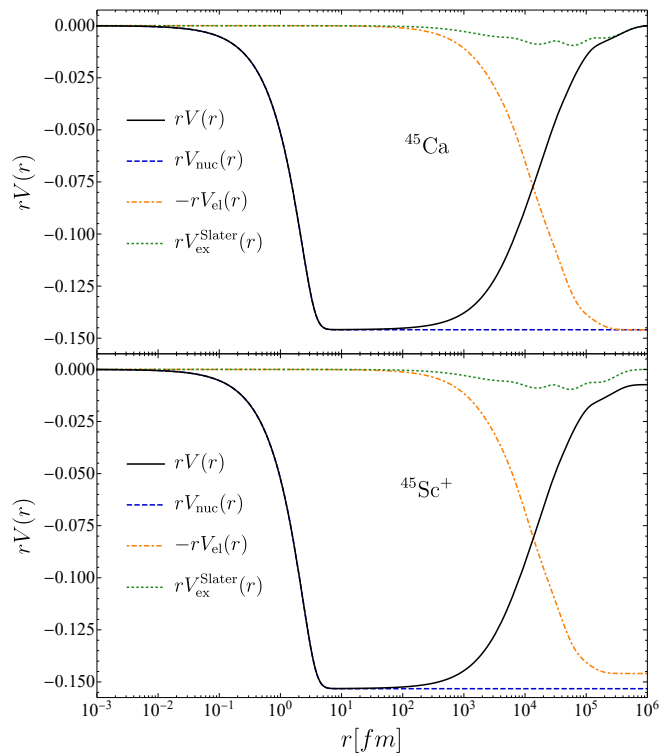}% Here is how to import EPS art
	\caption{\label{fig:ModifiedPotentials} Modified self-consistent potentials (solid curves) for the initial neutral atom, $^{45}$Ca (top), and the final positive ion, $^{45}\textrm{Sc}^+$ (bottom). See Eq.\ref{eq:newPotential} and text for details. In both cases, we indicate the nuclear potential (dashed curve), the electronic potential (dot-dashed curve) and the Slater's exchange potential (dotted curve).} 
\end{figure}

With the exchange potential, $V_{\text{ex}}^{\text{Slater}}(r)$, the obtained self-consistent potential does not respect the correct asymptotic behavior,
\begin{eqnarray}
\lim\limits_{r\rightarrow\infty}rV_{\text{DHFS}}(r)=-\alpha(Z_p-N_e+1),
\end{eqnarray}
because the exchange term cannot cancel the self-interaction term from the electronic potential. The drawback is solved with the introduction of the Latter's tail correction \cite{LatterPR1955} for the exchange potential,
\begin{eqnarray}
\label{eq:LatterTail}
V_\text{ex}(r)=
\begin{cases} 
V_{\text{ex}}^{\text{Slater}}(r) & r < r_{\text{Latter}}, \\\\
-\frac{\alpha(Z_p-N_e+1)}{r}-V_{\text{nuc}}(r)-V_{\text{el}}(r) & r\geq r_{\text{Latter}}. 
\end{cases}
\end{eqnarray}
The cutoff radius, $r_{\text{Latter}}$, is determined by solving the equation
\begin{eqnarray}
V_{\text{nuc}}(r)+V_{\text{el}}(r)+V_{\text{ex}}^{\text{Slater}}(r)=-\frac{\alpha(Z_p-N_e+1)}{r}.
\end{eqnarray}

The electron density, $\rho(r)$, is obtained self-consistently \cite{LibermanCPC1971,LibermanPR1965}. The procedure starts with an approximate electron density obtained from the Moli\`ere parametrization  of the Thomas-Fermi potential \cite{MoliereZNA1947}. Then, the electron density is renewed iteratively from the obtained bound orbitals, until neither the DHFS potential nor the binding energies change in a specified tolerance.

\begin{table}[t]%The best place to locate the table environment is directly after its first reference in text
	\caption{\label{tab:BindingEnergies}%
		Binding energies for neutral atom $^{45}$Ca in eV. In the first column, containing the occupied shells, we indicate with $*$ the relevant shells for the calculation of the exchange correction. In the second and third column, we present the true DHFS self-consistent method binding energies and the results obtained with the modified potential, respectively. In the last column, the experimental values taken from \cite{CarlsonBook1975}, are presented.  
	}
	\begin{ruledtabular}
		\begin{tabular}{cccc}
			\textrm{Orbital} $(n\ell_{j})$&
			\text{$\epsilon_e^{n\kappa}\textrm{(true)}$}&
			\text{$\epsilon_e^{n\kappa}\textrm{(modified)}$}&
			\text{$\epsilon_e^{n\kappa}\textrm{(exp)}$ \cite{CarlsonBook1975}}\\
			\colrule
			$*1s_{1/2}$ & -4015.1 &-4015.1&  $ -4041\pm2 $\\
			$*2s_{1/2}$ & -434.1 &-434.1&  $ -441\pm2 $\\
			$*2p_{1/2}$ & -359.1 & -359.1& $ -353\pm2 $\\
			$2p_{3/2}$ & -355.2 & -355.2& $ -349\pm2 $\\
			$*3s_{1/2}$ & -53.2 & -53.2& $ -46\pm2 $\\
			$*3p_{1/2}$ & -34.0 & -34.0& $ -28\pm2 $\\
			$3p_{3/2}$ & -33.6 & -33.6& $ -28\pm2 $\\
			$*4s_{1/2}$ & -5.45 & -5.08& $ -6.113\pm0.01 $\\
		\end{tabular}
	\end{ruledtabular}
\end{table}

For continuum states ($E_e>m_e$), the wave functions are not square integrable, so usually the normalization is done on the energy scale, i.e.
\begin{eqnarray}
\braket{\psi_{E_e\kappa}}{\psi_{E_e'\kappa}}=\delta(E_e-E_e').
\end{eqnarray}
The large- and small-component radial functions satisfy, for large values of $p_er$, the following asymptotic conditions
\begin{eqnarray}
&&\begin{Bmatrix}
g_{\kappa}(E_e,r)\\
f_{\kappa}(E_e,r)
\end{Bmatrix}\\
&&\sim \frac{1}{p_er}
\begin{Bmatrix}
\sqrt{\frac{E_e+m_e}{2E_e}} \sin (p_er-l\frac{\pi}{2}+\delta_{\kappa}-\eta \ln (2p_er))\\
\sqrt{\frac{E_e-m_e}{2E_e}} \cos (p_er-l\frac{\pi}{2}+\delta_{\kappa}-\eta \ln (2p_er))
\end{Bmatrix}\nonumber
\end{eqnarray}
where $\delta_{\kappa}$ is the phase shift. For calculating the wave functions corresponding to continuum states, we have employed the program RADIAL.f (included in \cite{SalvatCPC2019}), which can offer results accurate to up to 13 or 14 decimal figures.

The easiest way to ensure the orthogonality between continuum and bound states, e.g., $\braket{\psi'_{E_es}}{\psi'_{ns}}=0$, is to use the same potential in the calculation of the continuum and bound wave functions for the final positive ion atom. The procedure we used is that when obtaining the bound states, in the last iteration of the DHFS self-consistent method, we do not impose the Latter's tail correction for the exchange potential, Eq.~(\ref{eq:LatterTail}). In this way the potential for both continuum and bound states is written as,
\begin{eqnarray}
\label{eq:newPotential}
V(r)=V_{\text{nuc}}(r)+V_{\text{el}}(r)+V_{\text{ex}}^{\text{Slater}}(r),
\end{eqnarray}
where the electronic and exchange components are obtained from the last iterated electron density. In this way, we respect the correct asymptotic condition for a scattering potential, $\lim\limits_{r\rightarrow\infty}rV(r)=-\alpha(Z_p-N_e)$. In Fig.~\ref{fig:ModifiedPotentials}, we present the modified potentials for the initial neutral atom $^{45}$Ca, and the final positive ion $^{45}\textrm{Sc}^+$. In what follows, when we use the potential from Eq.~\ref{eq:newPotential}, for both continuum and bound states, we call the procedure a modified DHFS self-consistent method. Contrary, when the Latter's tail correction is imposed on the exchange potential, we call the procedure a true DHFS self-consistent method.    

We indeed deviate from the true DHFS self-consistent method, but it turns out that the modifications in the binding energies and the bound wave functions are negligible. In Table~\ref{tab:BindingEnergies}, we present the binding energies of the occupied states for the neutral atom $^{45}$Ca. The difference between the true self-consistent DHFS potential and the modified potential leads to a negligible difference just in the binding energy of the last occupied orbital, $4s_{1/2}$.

\iffalse
Regarding the continuum states, we included an exchange potential component in the scattering potential. 
\textbf{In \cite{RileyJCP1975} is considered incorrect to use $V_{\text{ex}}^{\text{Slater}}(r)$ for scattering potential, and in \cite{Mougeot-PRA2014,SalvatCPC2019} is completely avoided. }
\fi

\section{Results and Discussion}

\subsection{Exchange effect for $^{14}$C, $^{45}$Ca, $^{63}$Ni and $^{241}$Pu \\ $\beta$ decay}

\begin{table}[t]%The best place to locate the table environment is directly after its first reference in text
	\caption{\label{tab:NuclearData}%
		The relevant nuclear data for the isotopes considered in this paper. For each ground state to ground state $\beta$ transition we present the endpoint (second column) and the initial and final spin-parity states, $J_i^\pi$ and $J_f^\pi$ (third column). The Q-values of each $\beta$-decay are from Ref. \cite{WangCPC2017}  
	}
	\begin{ruledtabular}
		\begin{tabular}{ccc}
			Isotope&
			$Q$-value (keV) &
			$J_i^\pi,J_f^\pi$\\
			\colrule
			$^{10}$C & 156.476(4)  & $0^+,1^+$\\
			$^{45}$Ca & 259.7(7)  &$7/2^-,7/2^-$\\
			$^{63}$Ni & 66.977(15) & $1/2^-,3/2^-$\\
			$^{241}$Pu & 20.78 (17) & $5/2^+,5/2^-$\\
		\end{tabular}
	\end{ruledtabular}
\end{table}

\begin{figure}[h]
	\includegraphics[width=0.48\textwidth]{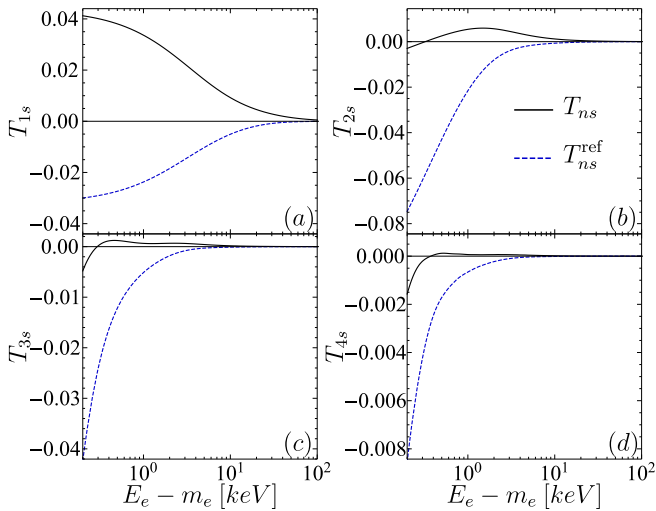}% Here is how to import EPS art
	\caption{The dimensionless quantities $T_{ns}$ (solid black) and $T^{\rm ref}_{ns}$ (dashed blue) defined in Eq.~\ref{eq:TnsQuantities} and ~\ref{eq:TrefnsQuantities}, respectively, necessary to perform the exchange correction calculation for the $\beta$-decay of $^{45}$Ca. The results are presented for $s_{1/2}$ orbitals with $n=1~(a)$, $n=2~(b)$, $n=3~(c)$, and $n=4~(d)$. The bound electron wave functions for both initial and final atoms are computed with the true DHFS method, and the continuum states of the final nucleus just with the electric and nuclear components of the potential.  \label{fig:badTQuantities}} 
\end{figure}

In order to investigate the steps of the exchange correction calculation, we selected four low Q-value $\beta$ transitions that were recently investigated, i.e., $^{14}$C \cite{WietfeldtPRC1995,LoidlJLTP2020}, $^{45}$Ca \cite{Hayen-RMP2018}, $^{63}$Ni \cite{Mougeot-PRA2014} and $^{241}$Pu \cite{Mougeot-PRA2012,Mougeot-PRA2014}. The relevant nuclear data for the considered isotopes are presented in Table~\ref{tab:NuclearData}. From the difference in the nuclear spin-parity states, the first three $\beta$ transitions are classified as allowed transitions and the last one as nonunique first forbidden, but it can be treated as allowed in the $\xi$ approximation \cite{Schopper-1966}. We perform the calculation of the exchange effect from $200$ eV up to the Q-value kinetic energy of the continuum electron for each transition (second column of Table~\ref{tab:NuclearData}).

To investigate the orthogonality effect on the exchange correction, we consider two approaches for computing the electron wave functions. In the first one, the bound wave functions for both the initial neutral atom and the final positive ion are obtained from a true DHFS self-consistent method. The continuum states for the emitted electron are obtained by solving Eq.~\ref{eq:radialEquations} with the nuclear and electronic potential of the final positive ion, i.e., $V(r)=V_{\text{nuc}}(r)+V_{\text{el}}(r)$. In the second approach, we consider the modified DHFS self-consistent method discussed in the previous section. For the bound and continuum electron states of the final positive ion, we use the same potential presented in Eq.~\ref{eq:newPotential} and depicted in the bottom part of Fig.~\ref{fig:ModifiedPotentials}.

\begin{figure}[h]
	\includegraphics[width=0.48\textwidth]{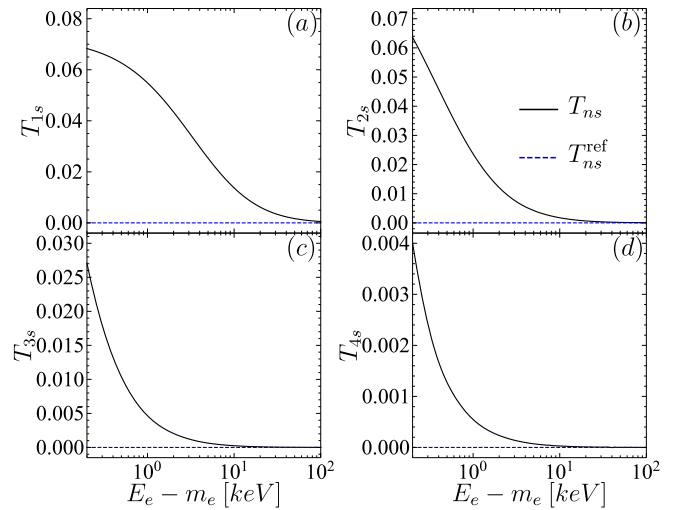}% Here is how to import EPS art
	\caption{The dimensionless quantities $T_{ns}$ (solid black) and $T^{\rm ref}_{ns}$ (dashed blue) defined in Eq.~\ref{eq:TnsQuantities} and ~\ref{eq:TrefnsQuantities}, respectively, necessary to perform the exchange correction calculation for the $\beta$-decay of $^{45}$Ca. The results are presented for $s_{1/2}$ orbitals with $n=1~(a)$, $n=2~(b)$, $n=3~(c)$, and $n=4~(d)$. The bound and continuum electron wave functions are computed with the modified DHFS self-consistent method. The zero value of the quantity $T^{\rm ref}_{ns}$ indicates a perfect orthogonality between the continuum states of the emitted electron and the bound states of the atomic electrons of the final nucleus. \label{fig:goodTQuantities}} 
\end{figure}

\begin{figure}[h]
	\includegraphics[width=0.4\textwidth]{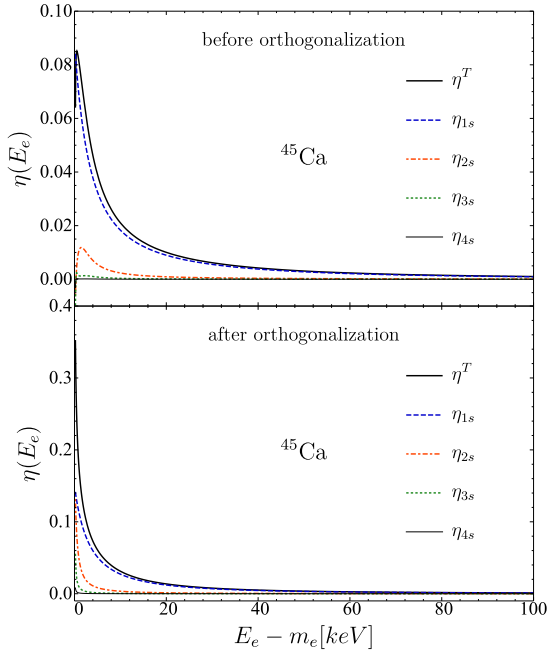}% Here is how to import EPS art
	\caption{The total exchange correction and the partial contributions from all occupied $s_{1/2}$ orbitals as functions of the kinetic energy of the electron emitted in the $\beta$-decay of $^{45}$Ca. The top figure is obtained with non-orthogonal continuum and bound states of the final atom (see text). In the bottom part the orthogonality is ensured by the modified DHFS self-consistent method.\label{fig:eta45Ca}} 
\end{figure}

The non-orthogonal bound and continuum states from the first approach lead to non-zero values of the dimensionless quantities $T^{\rm ref}_{ns}$, as can be seen in Fig.~\ref{fig:badTQuantities}, where we present all the contributions from the occupied $s_{1/2}$ orbitals in the case of $^{45}$Ca $\beta$-decay. In the second approach, where the orthogonality is imposed as descried above, one can see in Fig.~\ref{fig:goodTQuantities}, that $T^{\rm ref}_{ns}$ are constant and zero for any energy of the continuum state. Thus, the dimensionless quantities $T_{ns}$, entering the total exchange effect calculation, are strongly influenced by whether the orthogonality is imposed or not. We can say that as long as the overlaps $\braket{\psi'_{E_es}}{\psi'_{ns}}$ are not zero then the overlaps $\braket{\psi'_{E_es}}{\psi_{ns}}$, and implicitly the quantities $T_{ns}$, are in error.

For the $\beta$-decay of $^{45}$Ca, we display in Fig.~\ref{fig:eta45Ca} the total exchange effect, $\eta^T$, and the partial contributions, $\eta_{ns}$. The result with non-orthogonal states is presented in the top panel while the one with orthogonal states in the bottom panel. Not only that the exchange effect presents a completely different energy dependence, but the orthogonality of the wave functions strongly influences its magnitude. We mention that the differences in magnitude and energy dependence were pointed out also in \cite{Hayen-arxiv2020}, but a concrete explanation was not provided. In the first approach, the exchange effect has a maximum of $8\%$ at around $1$ keV, decreasing rapidly for lower energies of the continuum states. A similar exchange correction was obtained in \cite{Hayen-RMP2018} (see Fig.6, pg. 30), for $\beta$-decay of $^{45}$Ca. For orthogonal states, the exchange effect increases rapidly with decreasing energy of the continuum state and is about $35\%$ at $200$ eV. The downturn in the case of non-orthogonal states is associated with the behavior of the quantities $T_{ns}$, which are in error. From our investigation, the downturn can even change the sign of the exchange effect if the energy of the continuum state is low enough, which translates into a suppression of events in the low energy region of the $\beta$ spectrum. Therefore, the calculation of the exchange correction for ultra-low Q-value $\beta$ transitions should address the orthogonalization between the continuum and bound wave functions for the final atom. We mention that in the calculation of total exchange effect for $^{45}$Ca, we also included the contributions from the exchange with $p_{1/2}$ orbitals. Still, they are three orders of magnitude smaller than the $s_{1/2}$ contributions and can not be plotted in Fig.~\ref{fig:eta45Ca}. In the case of non-orthogonal states, the $p_{1/2}$ contributions also present a downturn, and they contribute to the downturn of the total exchange effect.

\begin{figure}[h]
	\includegraphics[width=0.48\textwidth]{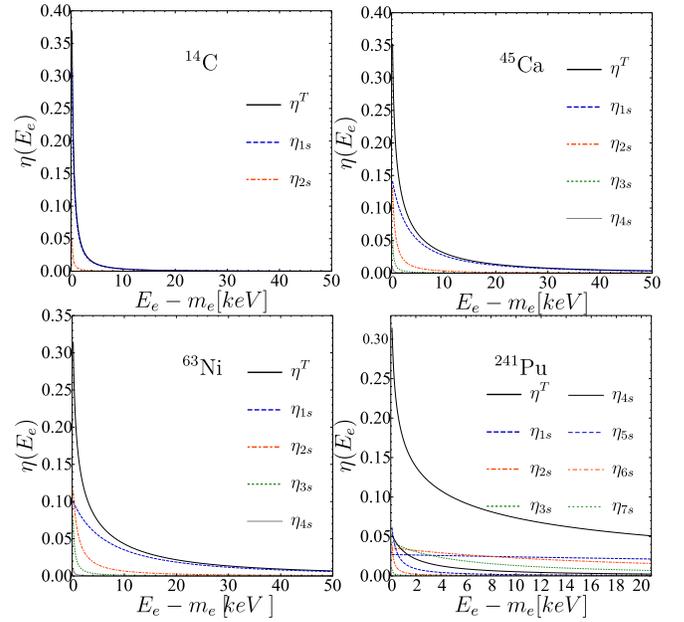}% Here is how to import EPS art
	\caption{The total exchange correction, $\eta^T$, and each partial contributions, $\eta_{ns}$, coming from the exchange with $s_{1/2}$ occupied orbitals as functions of the kinetic energy of the emitted electron, $E_e-m_e$. The results are presented for the $\beta$ decay of $^{14}$C, $^{45}$Ca, $^{63}$Ni and $^{241}$Pu. The total exchange correction also includes the partial contributions coming from the exchange with the $p_{1/2}$ occupied orbitals, which are too small to be included in the plot. \label{fig:multipleETA}} 
\end{figure}

\begin{figure}[h]
	\includegraphics[width=0.4\textwidth]{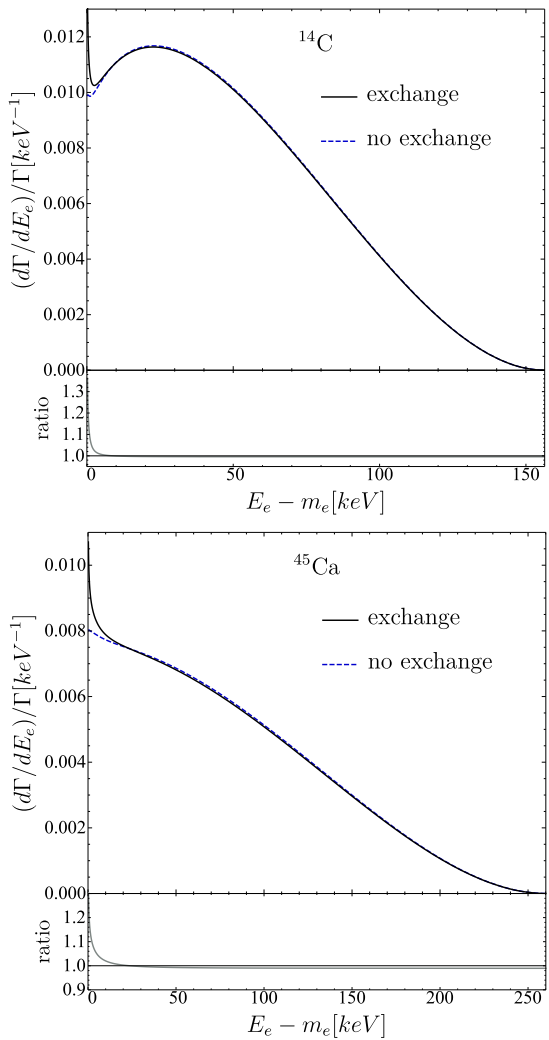}% Here is how to import EPS art
	\caption{The normalized electron spectra for the $\beta$ transitions of $^{14}$C and $^{45}$Ca, with the exchange effect included (solid black line) and without the exchange correction (dashed blue line). All spectra are normalized to unity over the full energy range. In the bottom panel of each $\beta$ emitter, we present the ratio between normalized spectrum with exchange correction and the normalized spectrum without the exchange correction.  \label{fig:elSpectra1}} 
\end{figure}

\begin{figure}[h]
	\includegraphics[width=0.4\textwidth]{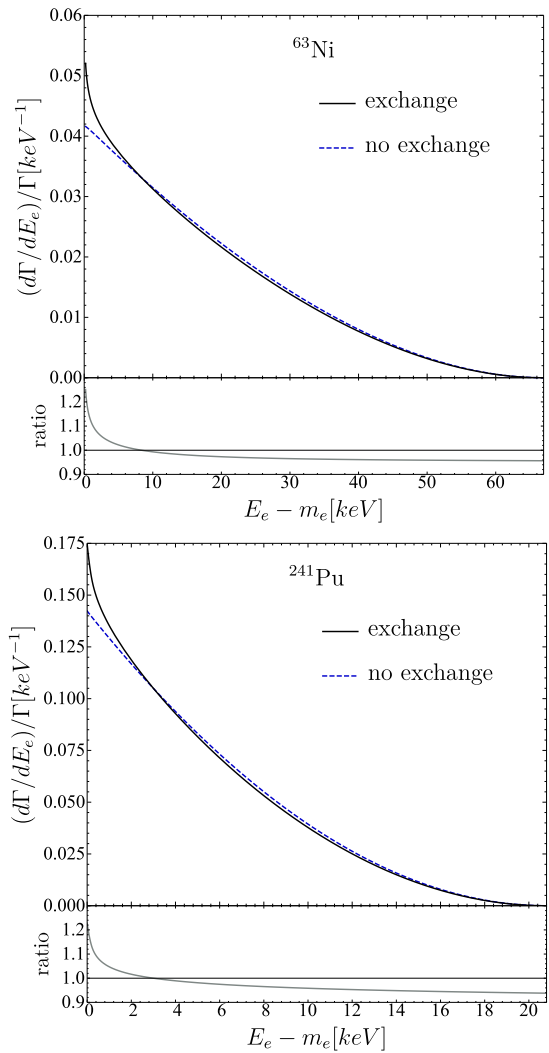}% Here is how to import EPS art
	\caption{Same as Fig.~\ref{fig:elSpectra1}, but for the $\beta$ transitions of  $^{63}$Ni and $^{241}$Pu.  \label{fig:elSpectra2}} 
\end{figure}

\begin{figure}[h]
	\includegraphics[width=0.48\textwidth]{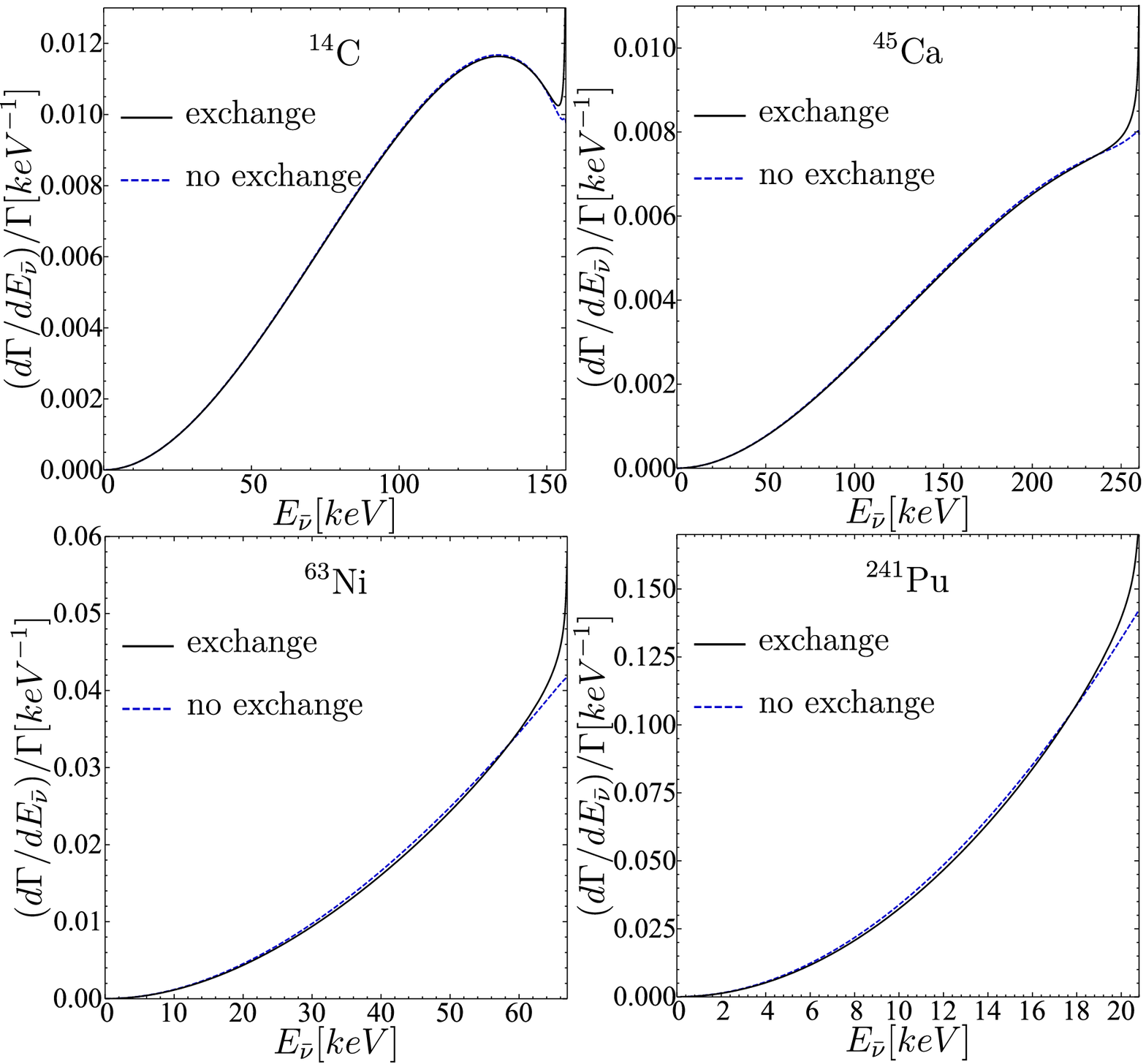}% Here is how to import EPS art
	\caption{The normalized antineutrino spectra corresponding to the $\beta$ transitions of $^{14}$C, $^{45}$Ca, $^{63}$Ni and $^{241}$Pu, with the exchange effect included (solid black line) and without the exchange correction (dashed blue line).\label{fig:nuSpectra}} 
\end{figure}

In the following, we present results for the ground state to ground state $\beta$ transitions of $^{14}$C, $^{45}$Ca, $^{63}$Ni and $^{241}$Pu, using the modified DHFS self-consistent method and the orthogonality between the continuum and bound states. In Fig.~\ref{fig:multipleETA}, the total exchange correction is presented for each transition, along with the contributions from the exchange with all occupied $s_{1/2}$ orbitals. The $p_{1/2}$ contributions are also included in the calculation, and they have similar behavior to the $s_{1/2}$ contributions, but are three orders of magnitude smaller than those ones. At the lowest energy chosen in this subsection for the continuum state, i.e. $200$ eV, the exchange corrections are about $37\%$, $35\%$, $32\%$ and $27\%$ for the $\beta$ decay of $^{14}$C, $^{45}$Ca, $^{63}$Ni and $^{241}$Pu, respectively. At first glance, the magnitude of the exchange effect decreases with the increasing nuclear charge, at least at $200$ eV. We will see in the following subsection that the magnitude of the exchange correction has a more complicated dependence on the atomic number as a result of a complex combined effect of the spatial extension of the atomic potential, the spatial extension of bound wave functions for different $n$ and the closure of $s_{1/2}$ and $p_{1/2}$ orbitals. From this study of one light, two light-medium, and one heavy nucleus, we can also see that the heavier the nucleus is, the higher energies the exchange correction affects. We will come back to this in the following subsection.

We present in Fig.~\ref{fig:elSpectra1} and Fig.~\ref{fig:elSpectra2} the normalized electron spectra corresponding to the decay of the four isotopes considered. For each transition, we display the spectrum with and without exchange correction with a solid and dashed line, respectively. Because of our choice of atomic potential, we mention that the spectra without the exchange effect already include the finite nuclear size, diffuse nuclear surface, and atomic screening corrections. The continuum wave functions for the emitted electrons, entering the Fermi functions from Eq.~\ref{eq:FermiFunction}, encode those corrections. We did not include in the calculations the radiative corrections, which are non-static Coulomb corrections. They are negligible for low $Q$-value $\beta$ transitions \cite{Mougeot-PRA2014,MougeotPRC2015}. The exchange effect dramatically increases the number of events in the low-energy region of the spectra and modifies its overall shape. In the case of non-orthogonal bound and continuum states for the final atom, the spectrum with the exchange effect presents a downturn, leading to a mismatch between the predicted spectrum and the measured one \cite{Mougeot-PRA2014,LoidlJLTP2020,KossertARI2022}. A comprehensive investigation of the residuals between the measured spectra and our result will be detailed elsewhere.

For completeness, we also show the corresponding antineutrino spectra of each transition in Fig.~\ref{fig:nuSpectra}. The antineutrino spectrum is obtained by the replacement, $E_e\rightarrow E_0-E_e$, in the electron spectrum from Eq.~\ref{eq:electronSpectrum}. The inversion is symmetric, and due to the exchange effect, we obtain the same spectrum shape modifications for antineutrino as for electrons. In the case of larger Q-value $\beta$ transitions, for which the radiative correction can not be neglected, the inversion is no longer symmetric because of the different expressions of the radiative corrections for electrons and antineutrinos. Important to notice is the fact that the exchange correction affects the endpoint regions of the antineutrino spectra. We leave the discussion open on how the exchange effect alters the cumulative antineutrino spectrum of a nuclear reactor. Further investigations are in progress in this direction.                          

\subsection{Analytical parametrization}

In order to provide an estimation of the exchange effects on a $\beta$-decay spectrum in general, we have performed the exchange correction calculations for a wide range of atomic numbers for the initial nucleus, from 1 to 102. The electrons emitted in the $\beta$ decay process were considered with kinetic energy between $50$ eV and $200$ keV.

\begin{figure}[t]
	\includegraphics[width=0.48\textwidth]{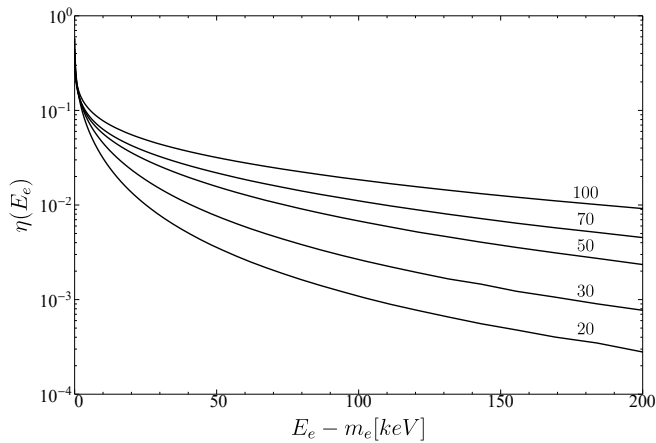}% Here is how to import EPS art
	\caption{The energy dependence of the total exchange correction for $\beta$ decay of different initial nuclei with atomic numbers $Z=20,30,50,70,100$. The quantities are computed using Eq.~\ref{eq:TotalEchangeCorrection}, which includes exchange with all occupied $s_{1/2}$ and $p_{1/2}$ orbitals.\label{fig:EtaEnergyDependence}} 
\end{figure}

We depict in Fig.~\ref{fig:EtaEnergyDependence} the energy dependence of the total exchange correction for the $\beta$ transitions  of different initial isotopes with $Z=20,30,50,70,100$. We included the contributions coming from the exchange with all occupied $s_{1/2}$ and $p_{1/2}$ orbitals. The decrease of the exchange effect with the increasing energy of the emitted electron is not a new or surprising result \cite{Harston-PRA1992}. Still, in comparison with the unscreened hydrogenic approximation used in the previous investigation \cite{Harston-PRA1992}, the DHFS self-consistent method employed here increases the overall magnitude of the exchange correction in the very low energy region. Another contribution to this difference is the fact that in \cite{Harston-PRA1992}, the calculations are performed including just the exchange with occupied $s_{1/2}$ orbitals up to and including the $3s_{1/2}$ orbital and completely neglecting the $p_{1/2}$ orbitals. Except for the low-energy region, the total exchange correction progressively increases with the nuclear charge. Still, even for $Z=100$, the total exchange effect is under $1\%$ at $200$ keV kinetic energy of the emitted electron. Its calculation can be neglected after this threshold, depending on the aimed accuracy.

\begin{figure*}[t]
	\includegraphics[width=0.8\textwidth]{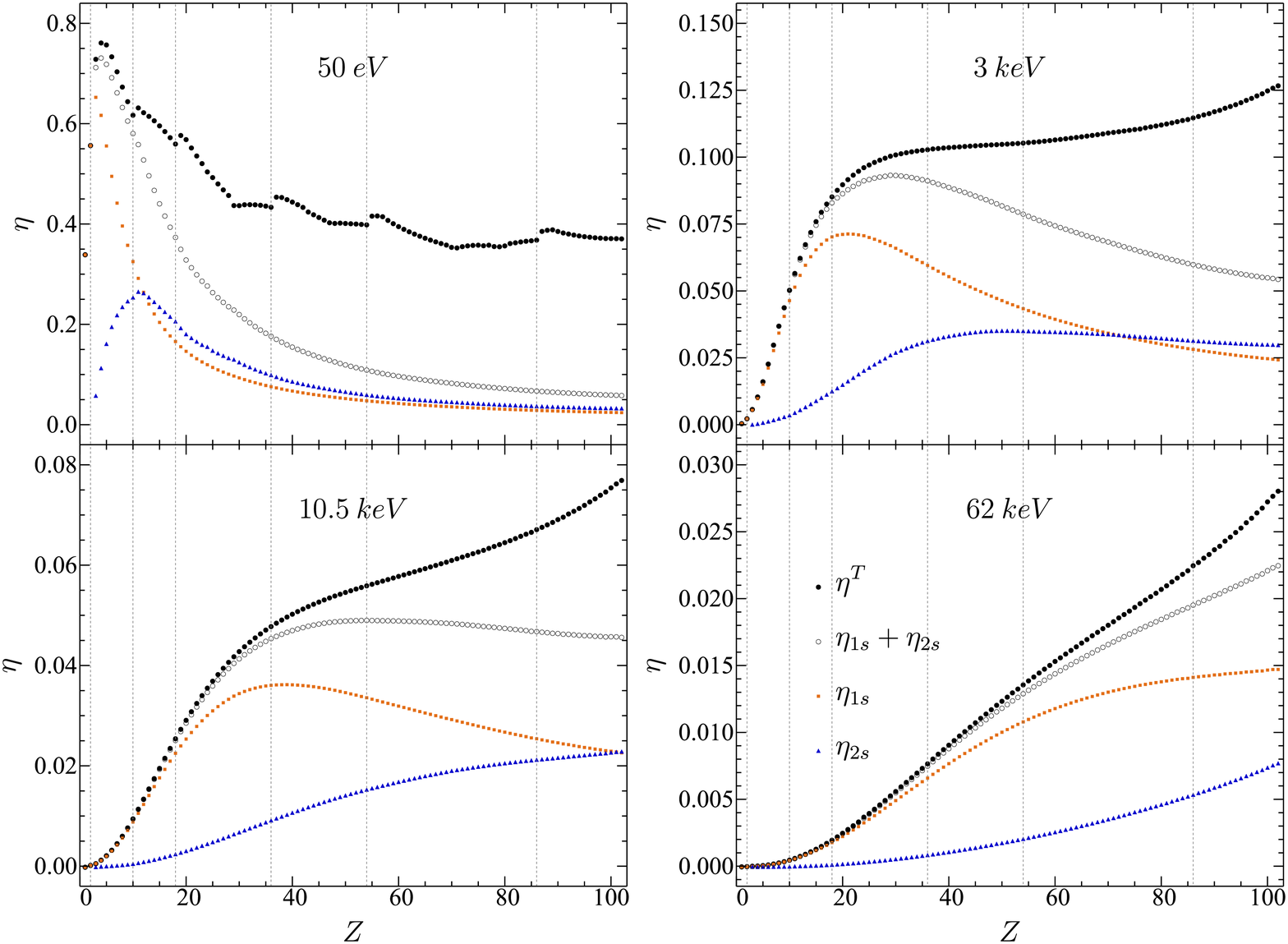}% Here is how to import EPS art
	\caption{The exchange correction as function of the atomic number of the $\beta$ emitter, from $Z=1$ to $Z=102$. The dependence is presented at four different kinetic energies of the $\beta$ electron. We depict the total exchange correction, $\eta^T$, with filled black circles and the partial contributions, $\eta_{1s}$ and $\eta_{2s}$, with filled orange squares and filled blue triangles, respectively. The empty black circles represent the sum of the partial contributions coming from the exchange with $1s_{1/2}$ and $2s_{1/2}$, i.e., $\eta_{1s}$+$\eta_{2s}$. We also indicate with thin dashed lines the atomic numbers where the $s_{1/2}$ orbitals are fully filled with bound electrons. \label{fig:etaZDependence}} 
\end{figure*}

We present in Fig.~\ref{fig:etaZDependence}, the exchange correction as a function of the atomic number of the $\beta$ emitter, from $Z=1$ to $Z=102$. The dependence is presented at four different kinetic energies of the $\beta$ electron, i.e., $50$ eV, $3$ keV, $10.5$ keV and $62$ keV. We depict the total exchange correction, $\eta^T$, with filled black circles and the partial contributions, $\eta_{1s}$ and $\eta_{2s}$, with filled orange squares and filled blue triangles, respectively. The empty black circles represent the sum of the partial contributions coming from the exchange with $1s_{1/2}$ and $2s_{1/2}$ orbitals, i.e., $\eta_{1s}$+$\eta_{2s}$. In this way, we can understand the importance of the contributions from the remaining occupied $s_{1/2}$ and $p_{1/2}$ orbitals.

At $5$ eV energy, the total exchange effect presents discontinuities. We can understand this behavior from the expression of $\eta_s$ in Eq.~\ref{eq:PartialExchnageSOrbitals}. With the increasing atomic number, new $s_{1/2}$ orbitals are filled with electrons, and new partial contributions enter in-game. Those contributions added after a shell closure are not zero in the low-energy region, which can explain the discontinuities in the total exchange effect. We indicate with thin dashed lines the atomic numbers where $s_{1/2}$ orbitals are filled with bound electrons, and we can see a correlation between the shell closure and the jumps in the total exchange correction. The $p_{1/2}$ orbitals closure also contributes to this behavior but subtly. With increasing energy, the jumps transform into a smooth dependence. Even for $3$ keV, the partial contributions start from zero, and the total exchange effect is a smooth function of the atomic number. It remains smooth with the increasing energy of the emitted electron. 

Regarding the partial contributions, $\eta_{1s}$ and $\eta_{2s}$, we point out a completely different $Z$ dependence than the one presented in \cite{Hayen-RMP2018} at $3$ keV. We associate the differences in Z dependence and the magnitude with the fact that the bound and continuum states of the final atom are orthogonal in our study. In our case, at $3$ keV, $\eta_{2s}$ presents a smooth dependence on the atomic number, but in \cite{Hayen-RMP2018} it jumps even on negative values for certain nuclear charges, leading to the downturn of the total exchange effect. From the sum of $\eta_{1s}$ and $\eta_{2s}$, we can see that, for really low energies, the higher partial contributions are essential for the calculation of the total exchange correction. Their importance decreases with the energy increasing for the continuum state.

The complex combined effect of the spatial extension of the atomic potential, the spatial extension of bound wave functions for different $n$ and the $s_{1/2}$ and $p_{1/2}$ shell closure makes the analytical parametrization of the total exchange correction difficult over the full $ E_e-m_e$ and $ Z $ range. We propose the following analytical fit for total exchange correction,
\begin{equation}
\label{eq:fit}
\eta^T(x)=\left(a+bx^c\right)\exp(-dx^e),
\end{equation}
as a function of the kinetic energy of the emitted electron, i.e., $x=E_e-m_e$ in keV. The jumps in the $Z$ dependence for low energies forced us to tabulate the required fit parameters for each $Z$ individually. The five fit parameters are given in Table~\ref{tab:fitParameters} (Supplemental Material). Because of the orthogonality imposed in our study, the fit is more relaxed than the one from \cite{Hayen-RMP2018}, which involves nine fit parameters. The discrepancy between the analytical fit models arises because the downturn in the exchange correction from \cite{Hayen-RMP2018} can not be approximated with a simple model. In our case, the difference between the fitted and calculated exchange effect never exceeded $10^{-3}$ over the full energy range tested, from $5$ eV to $200$ keV. Over the full $ E_e-m_e$ and $ Z $ range, the average of residuals is smaller than $10^{-4}$, and the fit is comparable in accuracy with the one from \cite{Hayen-RMP2018}. We mention that our fit is designed for extrapolation above $200$ keV kinetic energy of the emitted electron.

\LTcapwidth=0.47\textwidth

%\begingroup
\begin{longtable}[H]{cccccc}%The best place to locate the table environment is directly after its first reference in text
	\caption{(Supplemental Material) The fit parameters for the total exchange correction are tabulated individually for each atomic number of the initial nucleus. The dimension of each parameter is found from the fit equation, Eq.\ref{eq:fit}, with $x=E_e-m_e$ in keV.     
	}	
	\label{tab:fitParameters}\\
	%	\begin{ruledtabular}
	%		\begin{tabular}{cccccc}
	\hline \hline
	$Z$& $a$& $b$& $c$& $d$& $e$\\
	\colrule
	\endfirsthead
	\hline \hline
	$Z$& $a$& $b$& $c$& $d$& $e$\\
	\colrule
	\endhead
	1	& 40.878        & 41.395    & 1.0627       &9.6525        & 0.2310  \\
	2	& 11.440        & 13.268    & 1.2686       &7.3986        & 0.2962  \\
	3	& 5.5124        & 0.1003    & 0.2022       &5.2566        & 0.3183  \\
	4	& 2.8551        & 0.4427    & 1.7697       &4.1501        & 0.3819  \\
	5	& 2.4633        & 0.1412    & 1.7544       &3.5635        & 0.3688  \\
	6	& 2.1398        & 0.0317    & 1.9398       &3.1399        & 0.3583  \\
	7	& 1.9035        & 0.0043    & 2.2619       &2.8400        & 0.3482  \\
	8	& 1.8035        &-0.1029    & 0.6761       &2.6039        & 0.3244  \\
	9	& 4.4452        & 29.653    & 0.6901       &5.5130        & 0.2581  \\
	10	& 5.1785        & 39.584    & 0.7405       &5.7240        & 0.2470  \\
	11	& 4.3112        & 19.394    & 0.6981       &5.0458        & 0.2539  \\
	12	& 3.8227        & 14.001    & 0.6797       &4.7287        & 0.2549  \\
	13	& 3.8809        & 13.420    & 0.6866       &4.6739        & 0.2506  \\
	14	& 3.4108        & 9.6263    & 0.6827       &4.3715        & 0.2550  \\
	15	& 2.8742        & 6.3302    & 0.6779       &4.0086        & 0.2626  \\
	16	& 2.4395        & 4.2232    & 0.6750       &3.6740        & 0.2707  \\
	17	& 2.1242        & 2.9843    & 0.6725       &3.3994        & 0.2773  \\
	18	& 1.8963        & 2.2428    & 0.6691       &3.1818        & 0.2821  \\
	19	& 2.5709        & 4.0673    & 0.7067       &3.6467        & 0.2615  \\
	20	& 2.1091        & 2.4885    & 0.7326       &3.2774        & 0.2764  \\
	21	& 1.8382        & 1.8290    & 0.7168       &3.0469        & 0.2815  \\
	22	& 1.6569        & 1.4564    & 0.7006       &2.8800        & 0.2842  \\
	23	& 1.5267        & 1.2281    & 0.6830       &2.7557        & 0.2850  \\
	24	& 1.2780        & 0.8165    & 0.6507       &2.4783        & 0.2927  \\
	25	& 1.3602        & 1.0057    & 0.6422       &2.6026        & 0.2823  \\
	26	& 1.3086        & 0.9690    & 0.6203       &2.5654        & 0.2792  \\
	27	& 1.2748        & 0.9753    & 0.5974       &2.5548        & 0.2748  \\
	28	& 1.2568        & 1.0258    & 0.5753       &2.5714        & 0.2693  \\
	29	& 1.1136        & 0.8685    & 0.5282       &2.4317        & 0.2695  \\
	30	& 1.2723        & 1.2967    & 0.5380       &2.6960        & 0.2552  \\
	31	& 1.5426        & 2.0748    & 0.5547       &3.0430        & 0.2393  \\
	32	& 1.7034        & 2.5015    & 0.5692       &3.1979        & 0.2324  \\
	33	& 1.7962        & 2.6806    & 0.5814       &3.2649        & 0.2292  \\
	34	& 1.7945        & 2.5440    & 0.5913       &3.2375        & 0.2295  \\
	35	& 1.7239        & 2.2223    & 0.5996       &3.1466        & 0.2323  \\
	36	& 1.6218        & 1.8588    & 0.6071       &3.0246        & 0.2364  \\
	37	& 3.2275        & 6.6088    & 0.6445       &4.0640        & 0.2018  \\
	38	& 2.5377        & 3.9319    & 0.6563       &3.6498        & 0.2150  \\
	39	& 2.0575        & 2.5016    & 0.6589       &3.3026        & 0.2263  \\
	40	& 1.7775        & 1.7946    & 0.6630       &3.0608        & 0.2347  \\
	41	& 1.5791        & 1.3588    & 0.6678       &2.8673        & 0.2420  \\
	42	& 1.4287        & 1.0662    & 0.6727       &2.7055        & 0.2482  \\
	43	& 1.2091        & 0.7074    & 0.6834       &2.4424        & 0.2605  \\
	44	& 1.1287        & 0.5908    & 0.6884       &2.3352        & 0.2651  \\
	45	& 1.0633        & 0.5044    & 0.6915       &2.2439        & 0.2689  \\
	46	& 1.0102        & 0.4394    & 0.6921       &2.1667        & 0.2717  \\
	47	& 0.9644        & 0.3887    & 0.6921       &2.0988        & 0.2741  \\
	48	& 0.9834        & 0.4124    & 0.6813       &2.1319        & 0.2699  \\
	49	& 1.0483        & 0.4948    & 0.6587       &2.2341        & 0.2607  \\
	50	& 1.0811        & 0.5367    & 0.6463       &2.2829        & 0.2555  \\
	51	& 1.1207        & 0.5906    & 0.6343       &2.3406        & 0.2502  \\
	52	& 1.1568        & 0.6395    & 0.6250       &2.3902        & 0.2457  \\
	53	& 1.1788        & 0.6679    & 0.6207       &2.4189        & 0.2429  \\
	54	& 1.1859        & 0.6723    & 0.6192       &2.4260        & 0.2415  \\
	55	& 2.6675        & 3.8795    & 0.5667       &3.6830        & 0.1882  \\
	56	& 2.4590        & 3.1784    & 0.5810       &3.5353        & 0.1935  \\
	57	& 2.0825        & 2.2149    & 0.5909       &3.2648        & 0.2026  \\
	58	& 1.9646        & 2.0665    & 0.5747       &3.2011        & 0.2025  \\
	59	& 1.8991        & 2.0272    & 0.5593       &3.1752        & 0.2013  \\
	60	& 5.7069        &-4.4807    & 0.0445       &2.0158        & 0.0958  \\
	61	& 4.7160        &-3.6434    & 0.0472       &1.8830        & 0.1018  \\
	62	& 1.9100        & 2.4129    & 0.5209       &3.2733        & 0.1927  \\
	63	& 1.9756        & 2.7329    & 0.5116       &3.3596        & 0.1886  \\
	64	& 2.0675        & 3.1615    & 0.5048       &3.4651        & 0.1843  \\
	65	& 2.1927        & 3.7411    & 0.5007       &3.5924        & 0.1797  \\
	66	& 2.3524        & 4.4939    & 0.4989       &3.7361        & 0.1751  \\
	67	& 2.5481        & 5.4568    & 0.4998       &3.8932        & 0.1706  \\
	68	& 2.7706        & 6.6084    & 0.5026       &4.0523        & 0.1665  \\
	69	& 3.0281        & 8.0113    & 0.5070       &4.2161        & 0.1626  \\
	70	& 3.5968        & 12.385    & 0.5171       &4.5873        & 0.1551  \\
	71	& 3.3291        & 9.9229    & 0.5153       &4.4002        & 0.1584  \\
	72	& 2.9497        & 7.5192    & 0.5117       &4.1643        & 0.1631  \\
	73	& 2.5073        & 5.3125    & 0.5071       &3.8723        & 0.1695  \\
	74	& 2.1592        & 3.8373    & 0.5040       &3.6063        & 0.1760  \\
	75	& 1.9041        & 2.8882    & 0.5026       &3.3813        & 0.1819  \\
	76	& 2.1691        & 3.8474    & 0.5010       &3.6090        & 0.1746  \\
	77	& 0.0297        & 0.2031    &-0.2002       &0.3543        & 0.3930  \\
	78	& 1.2728        & 1.1235    & 0.5013       &2.6830        & 0.2041  \\
	79	&-0.1392        & 0.3445    &-0.1388       &0.2235        & 0.3614  \\
	80	&-0.1217        & 0.3229    &-0.1476       &0.2014        & 0.3780  \\
	81	&-0.0768        & 0.2711    &-0.1728       &0.1647        & 0.4341  \\
	82	&-0.0763        & 0.2694    &-0.1750       &0.1562        & 0.4366  \\
	83	&-0.0702        & 0.2624    &-0.1798       &0.1488        & 0.4456  \\
	84	&-0.0712        & 0.2625    &-0.1809       &0.1414        & 0.4469  \\
	85	&-0.0732        & 0.2638    &-0.1811       &0.1359        & 0.4454  \\
	86	& 1.4297        & 1.2203    & 0.5313       &2.7644        & 0.2004  \\
	87	& 52.470        &-39.955    & 0.0481       &4.3172        & 0.0384  \\
	88	& 3.4173        & 5.9999    & 0.5508       &4.0237        & 0.1635  \\
	89	& 0.1538        & 0.1231    &-0.3077       &0.4941        & 0.3504  \\
	90	& 0.0872        & 0.1567    &-0.2640       &0.3621        & 0.3864  \\
	91	& 0.0184        & 0.1954    &-0.2298       &0.2262        & 0.4372  \\
	92	&-0.0480        & 0.2417    &-0.2002       &0.1214        & 0.4769  \\
	93	&-0.0719        & 0.2662    &-0.1847       &0.1198        & 0.4457  \\
	94	& 5.5119        &-4.1559    & 0.0482       &2.0596        & 0.0783  \\
	95	& 4.6253        &-3.5213    & 0.0463       &1.8487        & 0.0863  \\
	96	&-0.1264        & 0.3340    &-0.1500       &0.1698        & 0.3550  \\
	97	& 3.3772        &-2.4923    & 0.0509       &1.6164        & 0.0989  \\
	98	& 3.0700        &-2.2689    & 0.0504       &1.5107        & 0.1049  \\
	99	& 2.7323        &-1.9692    & 0.0543       &1.4570        & 0.1093  \\
	100	& 1.2245        & 0.8824    & 0.4761       &2.4648        & 0.2000  \\
	101	& 1.2193        & 0.8939    & 0.4675       &2.4613        & 0.1988  \\
	102	& 1.2219        & 0.9252    & 0.4585       &2.4717        & 0.1971  \\
	
	\hline \hline
	
	%	\end{tabular}
	%		\end{ruledtabular}
\end{longtable} 
%\endgroup

\section{Conclusions}

The $\beta$-spectrum shape investigation is a powerful tool to answer and further understand open questions related to physics beyond the Standard Model or neutrino physics. With the increasing experimental capabilities of measuring the low-energy region, i.e., the first few keV, and the increasing statistics, an accurate theoretical prediction for the $\beta$-spectrum should be provided. Still, its description is a challenging task considering the multitude of effects that are important in the low-energy region. For allowed transitions, the primary atomic effects arise from the screened wave function of the emitted electron due to the atomic electron cloud and the exchange between continuum and bound electrons. The latter is the so-called exchange correction.

We investigate the exchange effect for the allowed $\beta$ transitions including all contributions from the occupied $s_{1/2}$ and $p_{1/2}$ orbitals. For the electron wave functions, we employ the Dirac-Hartree-Fock-Slater self-consistent method. We modified the last iteration of the self-consistent method to ensure the orthogonality between the continuum and bound electron states in the potential of the final atom. We found that orthogonality is an essential ingredient in calculating the exchange correction. Non-orthogonal continuum and bound states of the final atom lead to undesired errors in the overlaps between the initial atom's bound states and the final atom's continuum states which translates into a downturn in the total exchange correction. After including the orthogonality in the calculation of the exchange effect, we found considerable differences in magnitude and energy dependence when compared with previous investigations. We argue that our findings can solve the mismatch between the previous predictions and experimental measurements in the low-energy region.

Then, motivated to obtain an analytical parametrization for the exchange correction, we extended the calculation for a wide range of $\beta$ emitters, with Z ranging from $1$ to $102$. Except for the low-energy region, the total exchange correction progressively increases with the nuclear charge. We found that for ultra-low energy, i.e., $5$ eV, the $Z$ dependence of total exchange effect is affected by the closure of $s_{1/2}$ and $p_{1/2}$ orbitals. For higher energies, the exchange effect presents a smooth dependence on the nuclear charge, which is an entirely different result than previous investigations. We associate this behavior with the orthogonality between the continuum and the bound states of the final atom. We also show that the contributions from orbitals higher than the $2s_{1/2}$ orbital are essential for correctly calculating the total effect, especially for low energies. Finally, we provide an analytical expression of the total exchange correction for each atomic number for easy implementation in experimental investigations.

\section*{ACKNOWLEDGMENTS}

F. \v{S}. acknowledges support by the VEGA Grant Agency of the
Slovak Republic under Contract No. 1/0607/20 and by the
Ministry of Education, Youth and Sports of the Czech
Republic under the INAFYM Grant No. CZ.02.1.01/0.0/0.0/16\_019/0000766.\\

 S.S acknowledges support by a grant of the Romanian Ministry of Research, Innovation and Digitalization through the project PN19-030102-INCDFM.\\
 
 The figures for this article have been created using the SciDraw scientific figure preparation system \cite{SciDraw}.

%\newpage
\bibliography{myBib}% Produces the bibliography via BibTeX.

\providecommand{\noopsort}[1]{}\providecommand{\singleletter}[1]{#1}%
\begin{thebibliography}{43}%
\makeatletter
\providecommand \@ifxundefined [1]{%
 \@ifx{#1\undefined}
}%
\providecommand \@ifnum [1]{%
 \ifnum #1\expandafter \@firstoftwo
 \else \expandafter \@secondoftwo
 \fi
}%
\providecommand \@ifx [1]{%
 \ifx #1\expandafter \@firstoftwo
 \else \expandafter \@secondoftwo
 \fi
}%
\providecommand \natexlab [1]{#1}%
\providecommand \enquote  [1]{``#1''}%
\providecommand \bibnamefont  [1]{#1}%
\providecommand \bibfnamefont [1]{#1}%
\providecommand \citenamefont [1]{#1}%
\providecommand \href@noop [0]{\@secondoftwo}%
\providecommand \href [0]{\begingroup \@sanitize@url \@href}%
\providecommand \@href[1]{\@@startlink{#1}\@@href}%
\providecommand \@@href[1]{\endgroup#1\@@endlink}%
\providecommand \@sanitize@url [0]{\catcode `\\12\catcode `\$12\catcode
  `\&12\catcode `\#12\catcode `\^12\catcode `\_12\catcode `\%12\relax}%
\providecommand \@@startlink[1]{}%
\providecommand \@@endlink[0]{}%
\providecommand \url  [0]{\begingroup\@sanitize@url \@url }%
\providecommand \@url [1]{\endgroup\@href {#1}{\urlprefix }}%
\providecommand \urlprefix  [0]{URL }%
\providecommand \Eprint [0]{\href }%
\providecommand \doibase [0]{https://doi.org/}%
\providecommand \selectlanguage [0]{\@gobble}%
\providecommand \bibinfo  [0]{\@secondoftwo}%
\providecommand \bibfield  [0]{\@secondoftwo}%
\providecommand \translation [1]{[#1]}%
\providecommand \BibitemOpen [0]{}%
\providecommand \bibitemStop [0]{}%
\providecommand \bibitemNoStop [0]{.\EOS\space}%
\providecommand \EOS [0]{\spacefactor3000\relax}%
\providecommand \BibitemShut  [1]{\csname bibitem#1\endcsname}%
\let\auto@bib@innerbib\@empty
%</preamble>
\bibitem [{\citenamefont {Aker}\ \emph {et~al.}(2022)\citenamefont {Aker},
  \citenamefont {Beglarian}, \citenamefont {Behrens}, \citenamefont {Berlev},
  \citenamefont {Besserer}, \citenamefont {Bieringer} \emph
  {et~al.}}]{KATRIN-N2022}%
  \BibitemOpen
  \bibfield  {author} {\bibinfo {author} {\bibfnamefont {M.}~\bibnamefont
  {Aker}}, \bibinfo {author} {\bibfnamefont {A.}~\bibnamefont {Beglarian}},
  \bibinfo {author} {\bibfnamefont {J.}~\bibnamefont {Behrens}}, \bibinfo
  {author} {\bibfnamefont {A.}~\bibnamefont {Berlev}}, \bibinfo {author}
  {\bibfnamefont {U.}~\bibnamefont {Besserer}}, \bibinfo {author}
  {\bibfnamefont {F.}~\bibnamefont {Bieringer}, \bibfnamefont {B.~Block}},
  \emph {et~al.} (\bibinfo {collaboration} {The KATRIN Collaboration}),\
  }\bibfield  {title} {\bibinfo {title} {Direct neutrino-mass measurement with
  sub-electronvolt sensitivity},\ }\href
  {https://doi.org/10.1038/s41567-021-01463-1} {\bibfield  {journal} {\bibinfo
  {journal} {Nature Physics}\ }\textbf {\bibinfo {volume} {18}},\ \bibinfo
  {pages} {160} (\bibinfo {year} {2022})}\BibitemShut {NoStop}%
\bibitem [{\citenamefont {Herczeg}(2001)}]{HerczegPPNP2001}%
  \BibitemOpen
  \bibfield  {author} {\bibinfo {author} {\bibfnamefont {P.}~\bibnamefont
  {Herczeg}},\ }\bibfield  {title} {\bibinfo {title} {Beta decay beyond the
  standard model},\ }\href
  {https://doi.org/https://doi.org/10.1016/S0146-6410(01)00149-1} {\bibfield
  {journal} {\bibinfo  {journal} {Progress in Particle and Nuclear Physics}\
  }\textbf {\bibinfo {volume} {46}},\ \bibinfo {pages} {413} (\bibinfo {year}
  {2001})}\BibitemShut {NoStop}%
\bibitem [{\citenamefont {Nico}\ and\ \citenamefont
  {Snow}(2005)}]{NicoARNPS2005}%
  \BibitemOpen
  \bibfield  {author} {\bibinfo {author} {\bibfnamefont {J.~S.}\ \bibnamefont
  {Nico}}\ and\ \bibinfo {author} {\bibfnamefont {W.~M.}\ \bibnamefont
  {Snow}},\ }\bibfield  {title} {\bibinfo {title} {Fundamental neutron
  physics},\ }\href {https://doi.org/10.1146/annurev.nucl.55.090704.151611}
  {\bibfield  {journal} {\bibinfo  {journal} {Annual Review of Nuclear and
  Particle Science}\ }\textbf {\bibinfo {volume} {55}},\ \bibinfo {pages} {27}
  (\bibinfo {year} {2005})},\ \Eprint
  {https://arxiv.org/abs/https://doi.org/10.1146/annurev.nucl.55.090704.151611}
  {https://doi.org/10.1146/annurev.nucl.55.090704.151611} \BibitemShut
  {NoStop}%
\bibitem [{\citenamefont {Severijns}\ \emph {et~al.}(2006)\citenamefont
  {Severijns}, \citenamefont {Beck},\ and\ \citenamefont
  {Naviliat-Cuncic}}]{Severijns-RMP2006}%
  \BibitemOpen
  \bibfield  {author} {\bibinfo {author} {\bibfnamefont {N.}~\bibnamefont
  {Severijns}}, \bibinfo {author} {\bibfnamefont {M.}~\bibnamefont {Beck}},\
  and\ \bibinfo {author} {\bibfnamefont {O.}~\bibnamefont {Naviliat-Cuncic}},\
  }\bibfield  {title} {\bibinfo {title} {Tests of the standard electroweak
  model in nuclear beta decay},\ }\href
  {https://doi.org/10.1103/RevModPhys.78.991} {\bibfield  {journal} {\bibinfo
  {journal} {Rev. Mod. Phys.}\ }\textbf {\bibinfo {volume} {78}},\ \bibinfo
  {pages} {991} (\bibinfo {year} {2006})}\BibitemShut {NoStop}%
\bibitem [{\citenamefont {Severijns}\ and\ \citenamefont
  {Naviliat-Cuncic}(2011)}]{Severijns-ARNPS2011}%
  \BibitemOpen
  \bibfield  {author} {\bibinfo {author} {\bibfnamefont {N.}~\bibnamefont
  {Severijns}}\ and\ \bibinfo {author} {\bibfnamefont {O.}~\bibnamefont
  {Naviliat-Cuncic}},\ }\bibfield  {title} {\bibinfo {title} {Symmetry tests in
  nuclear beta decay},\ }\href
  {https://doi.org/10.1146/annurev-nucl-102010-130410} {\bibfield  {journal}
  {\bibinfo  {journal} {Annual Review of Nuclear and Particle Science}\
  }\textbf {\bibinfo {volume} {61}},\ \bibinfo {pages} {23} (\bibinfo {year}
  {2011})},\ \Eprint
  {https://arxiv.org/abs/https://doi.org/10.1146/annurev-nucl-102010-130410}
  {https://doi.org/10.1146/annurev-nucl-102010-130410} \BibitemShut {NoStop}%
\bibitem [{\citenamefont {Severijns}(2014)}]{Severijns-JPG2014}%
  \BibitemOpen
  \bibfield  {author} {\bibinfo {author} {\bibfnamefont {N.}~\bibnamefont
  {Severijns}},\ }\bibfield  {title} {\bibinfo {title} {Correlation and
  spectrum shape measurements in $\beta$-decay probing the standard model},\
  }\href {https://doi.org/10.1088/0954-3899/41/11/114006} {\bibfield  {journal}
  {\bibinfo  {journal} {Journal of Physics G: Nuclear and Particle Physics}\
  }\textbf {\bibinfo {volume} {41}},\ \bibinfo {pages} {114006} (\bibinfo
  {year} {2014})}\BibitemShut {NoStop}%
\bibitem [{\citenamefont {Hayen}\ \emph {et~al.}(2018)\citenamefont {Hayen},
  \citenamefont {Severijns}, \citenamefont {Bodek}, \citenamefont {Rozpedzik},\
  and\ \citenamefont {Mougeot}}]{Hayen-RMP2018}%
  \BibitemOpen
  \bibfield  {author} {\bibinfo {author} {\bibfnamefont {L.}~\bibnamefont
  {Hayen}}, \bibinfo {author} {\bibfnamefont {N.}~\bibnamefont {Severijns}},
  \bibinfo {author} {\bibfnamefont {K.}~\bibnamefont {Bodek}}, \bibinfo
  {author} {\bibfnamefont {D.}~\bibnamefont {Rozpedzik}},\ and\ \bibinfo
  {author} {\bibfnamefont {X.}~\bibnamefont {Mougeot}},\ }\bibfield  {title}
  {\bibinfo {title} {High precision analytical description of the allowed
  $\ensuremath{\beta}$ spectrum shape},\ }\href
  {https://doi.org/10.1103/RevModPhys.90.015008} {\bibfield  {journal}
  {\bibinfo  {journal} {Rev. Mod. Phys.}\ }\textbf {\bibinfo {volume} {90}},\
  \bibinfo {pages} {015008} (\bibinfo {year} {2018})}\BibitemShut {NoStop}%
\bibitem [{\citenamefont {Noordmans}\ \emph {et~al.}(2013)\citenamefont
  {Noordmans}, \citenamefont {Wilschut},\ and\ \citenamefont
  {Timmermans}}]{Noordmans-PRC2015}%
  \BibitemOpen
  \bibfield  {author} {\bibinfo {author} {\bibfnamefont {J.~P.}\ \bibnamefont
  {Noordmans}}, \bibinfo {author} {\bibfnamefont {H.~W.}\ \bibnamefont
  {Wilschut}},\ and\ \bibinfo {author} {\bibfnamefont {R.~G.~E.}\ \bibnamefont
  {Timmermans}},\ }\bibfield  {title} {\bibinfo {title} {Lorentz violation in
  neutron decay and allowed nuclear $\ensuremath{\beta}$ decay},\ }\href
  {https://doi.org/10.1103/PhysRevC.87.055502} {\bibfield  {journal} {\bibinfo
  {journal} {Phys. Rev. C}\ }\textbf {\bibinfo {volume} {87}},\ \bibinfo
  {pages} {055502} (\bibinfo {year} {2013})}\BibitemShut {NoStop}%
\bibitem [{\citenamefont {Bahcall}(1963)}]{Bahcall-PR1963}%
  \BibitemOpen
  \bibfield  {author} {\bibinfo {author} {\bibfnamefont {J.~N.}\ \bibnamefont
  {Bahcall}},\ }\bibfield  {title} {\bibinfo {title} {Overlap and exchange
  effects in beta decay},\ }\href {https://doi.org/10.1103/PhysRev.129.2683}
  {\bibfield  {journal} {\bibinfo  {journal} {Phys. Rev.}\ }\textbf {\bibinfo
  {volume} {129}},\ \bibinfo {pages} {2683} (\bibinfo {year}
  {1963})}\BibitemShut {NoStop}%
\bibitem [{\citenamefont {Haxton}(1985)}]{Haxton-PRL1985}%
  \BibitemOpen
  \bibfield  {author} {\bibinfo {author} {\bibfnamefont {W.~C.}\ \bibnamefont
  {Haxton}},\ }\bibfield  {title} {\bibinfo {title} {Atomic effects and
  heavy-neutrino emission in beta decay},\ }\href
  {https://doi.org/10.1103/PhysRevLett.55.807} {\bibfield  {journal} {\bibinfo
  {journal} {Phys. Rev. Lett.}\ }\textbf {\bibinfo {volume} {55}},\ \bibinfo
  {pages} {807} (\bibinfo {year} {1985})}\BibitemShut {NoStop}%
\bibitem [{\citenamefont {Harston}\ and\ \citenamefont
  {Pyper}(1992)}]{Harston-PRA1992}%
  \BibitemOpen
  \bibfield  {author} {\bibinfo {author} {\bibfnamefont {M.~R.}\ \bibnamefont
  {Harston}}\ and\ \bibinfo {author} {\bibfnamefont {N.~C.}\ \bibnamefont
  {Pyper}},\ }\bibfield  {title} {\bibinfo {title} {Exchange effects in
  \ensuremath{\beta} decays of many-electron atoms},\ }\href
  {https://doi.org/10.1103/PhysRevA.45.6282} {\bibfield  {journal} {\bibinfo
  {journal} {Phys. Rev. A}\ }\textbf {\bibinfo {volume} {45}},\ \bibinfo
  {pages} {6282} (\bibinfo {year} {1992})}\BibitemShut {NoStop}%
\bibitem [{\citenamefont {Mougeot}\ \emph {et~al.}(2012)\citenamefont
  {Mougeot}, \citenamefont {B\'e}, \citenamefont {Bisch},\ and\ \citenamefont
  {Loidl}}]{Mougeot-PRA2012}%
  \BibitemOpen
  \bibfield  {author} {\bibinfo {author} {\bibfnamefont {X.}~\bibnamefont
  {Mougeot}}, \bibinfo {author} {\bibfnamefont {M.-M.}\ \bibnamefont {B\'e}},
  \bibinfo {author} {\bibfnamefont {C.}~\bibnamefont {Bisch}},\ and\ \bibinfo
  {author} {\bibfnamefont {M.}~\bibnamefont {Loidl}},\ }\bibfield  {title}
  {\bibinfo {title} {Evidence for the exchange effect in the
  $\ensuremath{\beta}$ decay of ${}^{241}${P}u},\ }\href
  {https://doi.org/10.1103/PhysRevA.86.042506} {\bibfield  {journal} {\bibinfo
  {journal} {Phys. Rev. A}\ }\textbf {\bibinfo {volume} {86}},\ \bibinfo
  {pages} {042506} (\bibinfo {year} {2012})}\BibitemShut {NoStop}%
\bibitem [{\citenamefont {Rose}(1961{\natexlab{a}})}]{Rose-1961}%
  \BibitemOpen
  \bibfield  {author} {\bibinfo {author} {\bibfnamefont {M.~E.}\ \bibnamefont
  {Rose}},\ }\href@noop {} {\emph {\bibinfo {title} {Relativistic electron
  theory}}}\ (\bibinfo  {publisher} {John Wiley and Sons},\ \bibinfo {year}
  {1961})\BibitemShut {NoStop}%
\bibitem [{\citenamefont {Mougeot}\ and\ \citenamefont
  {Bisch}(2014)}]{Mougeot-PRA2014}%
  \BibitemOpen
  \bibfield  {author} {\bibinfo {author} {\bibfnamefont {X.}~\bibnamefont
  {Mougeot}}\ and\ \bibinfo {author} {\bibfnamefont {C.}~\bibnamefont
  {Bisch}},\ }\bibfield  {title} {\bibinfo {title} {Consistent calculation of
  the screening and exchange effects in allowed
  ${\ensuremath{\beta}}^{\ensuremath{-}}$ transitions},\ }\href
  {https://doi.org/10.1103/PhysRevA.90.012501} {\bibfield  {journal} {\bibinfo
  {journal} {Phys. Rev. A}\ }\textbf {\bibinfo {volume} {90}},\ \bibinfo
  {pages} {012501} (\bibinfo {year} {2014})}\BibitemShut {NoStop}%
\bibitem [{\citenamefont {Akerib}\ \emph {et~al.}(2020)\citenamefont {Akerib}
  \emph {et~al.}}]{LUX-ZEPLINPRD2020}%
  \BibitemOpen
  \bibfield  {author} {\bibinfo {author} {\bibfnamefont {D.~S.}\ \bibnamefont
  {Akerib}} \emph {et~al.} (\bibinfo {collaboration} {LUX-ZEPLIN
  Collaboration}),\ }\bibfield  {title} {\bibinfo {title} {Projected wimp
  sensitivity of the lux-zeplin dark matter experiment},\ }\href
  {https://doi.org/10.1103/PhysRevD.101.052002} {\bibfield  {journal} {\bibinfo
   {journal} {Phys. Rev. D}\ }\textbf {\bibinfo {volume} {101}},\ \bibinfo
  {pages} {052002} (\bibinfo {year} {2020})}\BibitemShut {NoStop}%
\bibitem [{\citenamefont {Aprile}\ \emph
  {et~al.}(2020{\natexlab{a}})\citenamefont {Aprile} \emph
  {et~al.}}]{AprileJCAP2020}%
  \BibitemOpen
  \bibfield  {author} {\bibinfo {author} {\bibfnamefont {E.}~\bibnamefont
  {Aprile}} \emph {et~al.} (\bibinfo {collaboration} {XENON Collaboration}),\
  }\bibfield  {title} {\bibinfo {title} {Projected {WIMP} sensitivity of the
  {XENONnT} dark matter experiment},\ }\href
  {https://doi.org/10.1088/1475-7516/2020/11/031} {\bibfield  {journal}
  {\bibinfo  {journal} {Journal of Cosmology and Astroparticle Physics}\
  }\textbf {\bibinfo {volume} {2020}}\bibinfo  {number} { (11)},\ \bibinfo
  {pages} {031}}\BibitemShut {NoStop}%
\bibitem [{\citenamefont {Aprile}\ \emph
  {et~al.}(2020{\natexlab{b}})\citenamefont {Aprile} \emph
  {et~al.}}]{XENON-PRD2020}%
  \BibitemOpen
\bibfield  {number} {  }\bibfield  {author} {\bibinfo {author} {\bibfnamefont
  {E.}~\bibnamefont {Aprile}} \emph {et~al.} (\bibinfo {collaboration} {XENON
  Collaboration}),\ }\bibfield  {title} {\bibinfo {title} {Excess electronic
  recoil events in {XENON1T}},\ }\href
  {https://doi.org/10.1103/PhysRevD.102.072004} {\bibfield  {journal} {\bibinfo
   {journal} {Phys. Rev. D}\ }\textbf {\bibinfo {volume} {102}},\ \bibinfo
  {pages} {072004} (\bibinfo {year} {2020}{\natexlab{b}})}\BibitemShut
  {NoStop}%
\bibitem [{\citenamefont {Haselschwardt}\ \emph {et~al.}(2020)\citenamefont
  {Haselschwardt}, \citenamefont {Kostensalo}, \citenamefont {Mougeot},\ and\
  \citenamefont {Suhonen}}]{Haselschwardt-PRC2020}%
  \BibitemOpen
  \bibfield  {author} {\bibinfo {author} {\bibfnamefont {S.~J.}\ \bibnamefont
  {Haselschwardt}}, \bibinfo {author} {\bibfnamefont {J.}~\bibnamefont
  {Kostensalo}}, \bibinfo {author} {\bibfnamefont {X.}~\bibnamefont
  {Mougeot}},\ and\ \bibinfo {author} {\bibfnamefont {J.}~\bibnamefont
  {Suhonen}},\ }\bibfield  {title} {\bibinfo {title} {Improved calculations of
  $\ensuremath{\beta}$ decay backgrounds to new physics in liquid xenon
  detectors},\ }\href {https://doi.org/10.1103/PhysRevC.102.065501} {\bibfield
  {journal} {\bibinfo  {journal} {Phys. Rev. C}\ }\textbf {\bibinfo {volume}
  {102}},\ \bibinfo {pages} {065501} (\bibinfo {year} {2020})}\BibitemShut
  {NoStop}%
\bibitem [{\citenamefont {Loidl}\ \emph {et~al.}(2014)\citenamefont {Loidl},
  \citenamefont {Rodrigues}, \citenamefont {Le-Bret},\ and\ \citenamefont
  {Mougeot}}]{LoidlARI2014}%
  \BibitemOpen
  \bibfield  {author} {\bibinfo {author} {\bibfnamefont {M.}~\bibnamefont
  {Loidl}}, \bibinfo {author} {\bibfnamefont {M.}~\bibnamefont {Rodrigues}},
  \bibinfo {author} {\bibfnamefont {C.}~\bibnamefont {Le-Bret}},\ and\ \bibinfo
  {author} {\bibfnamefont {X.}~\bibnamefont {Mougeot}},\ }\bibfield  {title}
  {\bibinfo {title} {Beta spectrometry with metallic magnetic calorimeters},\
  }\href {https://doi.org/https://doi.org/10.1016/j.apradiso.2013.11.024}
  {\bibfield  {journal} {\bibinfo  {journal} {Applied Radiation and Isotopes}\
  }\textbf {\bibinfo {volume} {87}},\ \bibinfo {pages} {302} (\bibinfo {year}
  {2014})},\ \bibinfo {note} {proceedings of the 19th International Conference
  on Radionuclide Metrology and its Applications 17–21 June 2013, Antwerp,
  Belgium}\BibitemShut {NoStop}%
\bibitem [{\citenamefont {Loidl}\ \emph {et~al.}(2019)\citenamefont {Loidl},
  \citenamefont {Beyer}, \citenamefont {Bockhorn}, \citenamefont {Enss},
  \citenamefont {Kempf}, \citenamefont {Kossert}, \citenamefont {Mariam},
  \citenamefont {Nähle}, \citenamefont {Paulsen}, \citenamefont {Ranitzsch},
  \citenamefont {Rodrigues},\ and\ \citenamefont {Schmidt}}]{LoidlARI2019}%
  \BibitemOpen
  \bibfield  {author} {\bibinfo {author} {\bibfnamefont {M.}~\bibnamefont
  {Loidl}}, \bibinfo {author} {\bibfnamefont {J.}~\bibnamefont {Beyer}},
  \bibinfo {author} {\bibfnamefont {L.}~\bibnamefont {Bockhorn}}, \bibinfo
  {author} {\bibfnamefont {C.}~\bibnamefont {Enss}}, \bibinfo {author}
  {\bibfnamefont {S.}~\bibnamefont {Kempf}}, \bibinfo {author} {\bibfnamefont
  {K.}~\bibnamefont {Kossert}}, \bibinfo {author} {\bibfnamefont
  {R.}~\bibnamefont {Mariam}}, \bibinfo {author} {\bibfnamefont
  {O.}~\bibnamefont {Nähle}}, \bibinfo {author} {\bibfnamefont
  {M.}~\bibnamefont {Paulsen}}, \bibinfo {author} {\bibfnamefont
  {P.}~\bibnamefont {Ranitzsch}}, \bibinfo {author} {\bibfnamefont
  {M.}~\bibnamefont {Rodrigues}},\ and\ \bibinfo {author} {\bibfnamefont
  {M.}~\bibnamefont {Schmidt}},\ }\bibfield  {title} {\bibinfo {title} {Beta
  spectrometry with metallic magnetic calorimeters in the framework of the
  european empir project metrobeta},\ }\href
  {https://doi.org/https://doi.org/10.1016/j.apradiso.2019.108830} {\bibfield
  {journal} {\bibinfo  {journal} {Applied Radiation and Isotopes}\ }\textbf
  {\bibinfo {volume} {153}},\ \bibinfo {pages} {108830} (\bibinfo {year}
  {2019})}\BibitemShut {NoStop}%
\bibitem [{\citenamefont {Kossert}\ \emph {et~al.}(2022)\citenamefont
  {Kossert}, \citenamefont {Loidl}, \citenamefont {Mougeot}, \citenamefont
  {Paulsen}, \citenamefont {Ranitzsch},\ and\ \citenamefont
  {Rodrigues}}]{KossertARI2022}%
  \BibitemOpen
  \bibfield  {author} {\bibinfo {author} {\bibfnamefont {K.}~\bibnamefont
  {Kossert}}, \bibinfo {author} {\bibfnamefont {M.}~\bibnamefont {Loidl}},
  \bibinfo {author} {\bibfnamefont {X.}~\bibnamefont {Mougeot}}, \bibinfo
  {author} {\bibfnamefont {M.}~\bibnamefont {Paulsen}}, \bibinfo {author}
  {\bibfnamefont {P.}~\bibnamefont {Ranitzsch}},\ and\ \bibinfo {author}
  {\bibfnamefont {M.}~\bibnamefont {Rodrigues}},\ }\bibfield  {title} {\bibinfo
  {title} {High precision measurement of the $^{151}${S}m beta decay by means
  of a metallic magnetic calorimeter},\ }\href
  {https://doi.org/https://doi.org/10.1016/j.apradiso.2022.110237} {\bibfield
  {journal} {\bibinfo  {journal} {Applied Radiation and Isotopes}\ }\textbf
  {\bibinfo {volume} {185}},\ \bibinfo {pages} {110237} (\bibinfo {year}
  {2022})}\BibitemShut {NoStop}%
\bibitem [{\citenamefont {Behrens}\ and\ \citenamefont
  {B\"uring}(1982)}]{BehrensBook1982}%
  \BibitemOpen
  \bibfield  {author} {\bibinfo {author} {\bibfnamefont {H.}~\bibnamefont
  {Behrens}}\ and\ \bibinfo {author} {\bibfnamefont {W.}~\bibnamefont
  {B\"uring}},\ }\href@noop {} {\emph {\bibinfo {title} {Electron Radial Wave
  Functions and Nuclear Beta Decay}}}\ (\bibinfo  {publisher} {Clarendon
  Press},\ \bibinfo {year} {1982})\BibitemShut {NoStop}%
\bibitem [{\citenamefont {Doi}\ \emph {et~al.}(1985)\citenamefont {Doi},
  \citenamefont {Kotani},\ and\ \citenamefont {Takasugi}}]{DoiPTPS1985}%
  \BibitemOpen
  \bibfield  {author} {\bibinfo {author} {\bibfnamefont {M.}~\bibnamefont
  {Doi}}, \bibinfo {author} {\bibfnamefont {T.}~\bibnamefont {Kotani}},\ and\
  \bibinfo {author} {\bibfnamefont {E.}~\bibnamefont {Takasugi}},\ }\bibfield
  {title} {\bibinfo {title} {{Double Beta Decay and Majorana Neutrino}},\
  }\href {https://doi.org/10.1143/PTPS.83.1} {\bibfield  {journal} {\bibinfo
  {journal} {Progress of Theoretical Physics Supplement}\ }\textbf {\bibinfo
  {volume} {83}},\ \bibinfo {pages} {1} (\bibinfo {year} {1985})},\ \Eprint
  {https://arxiv.org/abs/https://academic.oup.com/ptps/article-pdf/doi/10.1143/PTPS.83.1/5227078/83-1.pdf}
  {https://academic.oup.com/ptps/article-pdf/doi/10.1143/PTPS.83.1/5227078/83-1.pdf}
  \BibitemShut {NoStop}%
\bibitem [{\citenamefont {Johansson}\ \emph {et~al.}(2000)\citenamefont
  {Johansson}, \citenamefont {Sherif},\ and\ \citenamefont
  {Ghoddoussi}}]{JohanssonNPA2000}%
  \BibitemOpen
  \bibfield  {author} {\bibinfo {author} {\bibfnamefont {J.}~\bibnamefont
  {Johansson}}, \bibinfo {author} {\bibfnamefont {H.}~\bibnamefont {Sherif}},\
  and\ \bibinfo {author} {\bibfnamefont {F.}~\bibnamefont {Ghoddoussi}},\
  }\bibfield  {title} {\bibinfo {title} {Orthogonality effects in relativistic
  models of nucleon knockout reactions},\ }\href
  {https://doi.org/https://doi.org/10.1016/S0375-9474(99)00434-0} {\bibfield
  {journal} {\bibinfo  {journal} {Nuclear Physics A}\ }\textbf {\bibinfo
  {volume} {665}},\ \bibinfo {pages} {403} (\bibinfo {year}
  {2000})}\BibitemShut {NoStop}%
\bibitem [{\citenamefont {Rose}(1961{\natexlab{b}})}]{RoseBook1961}%
  \BibitemOpen
  \bibfield  {author} {\bibinfo {author} {\bibfnamefont {M.~E.}\ \bibnamefont
  {Rose}},\ }\href@noop {} {\emph {\bibinfo {title} {Relativistic {E}lectron
  {T}heory}}}\ (\bibinfo  {publisher} {John Wiley and Sons},\ \bibinfo {year}
  {1961})\BibitemShut {NoStop}%
\bibitem [{\citenamefont {Rose}(1995)}]{RoseBook1995}%
  \BibitemOpen
  \bibfield  {author} {\bibinfo {author} {\bibfnamefont {M.~E.}\ \bibnamefont
  {Rose}},\ }\href@noop {} {\emph {\bibinfo {title} {Elementary {T}heory of
  {A}ngular {M}omentum}}}\ (\bibinfo  {publisher} {Dover},\ \bibinfo {year}
  {1995})\BibitemShut {NoStop}%
\bibitem [{\citenamefont {Varshalovich}\ \emph {et~al.}(1995)\citenamefont
  {Varshalovich}, \citenamefont {Moskalev},\ and\ \citenamefont
  {Khersonskii}}]{VarshalovichBook1988}%
  \BibitemOpen
  \bibfield  {author} {\bibinfo {author} {\bibfnamefont {D.~A.}\ \bibnamefont
  {Varshalovich}}, \bibinfo {author} {\bibfnamefont {A.~N.}\ \bibnamefont
  {Moskalev}},\ and\ \bibinfo {author} {\bibfnamefont {V.~K.}\ \bibnamefont
  {Khersonskii}},\ }\href@noop {} {\emph {\bibinfo {title} {Quantum {T}heory of
  {A}ngular {M}omentum}}}\ (\bibinfo  {publisher} {World Scientific},\ \bibinfo
  {year} {1995})\BibitemShut {NoStop}%
\bibitem [{\citenamefont {Salvat}\ and\ \citenamefont
  {Fernández-Varea}(2019)}]{SalvatCPC2019}%
  \BibitemOpen
  \bibfield  {author} {\bibinfo {author} {\bibfnamefont {F.}~\bibnamefont
  {Salvat}}\ and\ \bibinfo {author} {\bibfnamefont {J.~M.}\ \bibnamefont
  {Fernández-Varea}},\ }\bibfield  {title} {\bibinfo {title} {{RADIAL}: A
  fortran subroutine package for the solution of the radial {S}chr\"odinger and
  {D}irac wave equations},\ }\href
  {https://doi.org/https://doi.org/10.1016/j.cpc.2019.02.011} {\bibfield
  {journal} {\bibinfo  {journal} {Computer Physics Communications}\ }\textbf
  {\bibinfo {volume} {240}},\ \bibinfo {pages} {165} (\bibinfo {year}
  {2019})}\BibitemShut {NoStop}%
\bibitem [{\citenamefont {Slater}(1951)}]{SlaterPR1951}%
  \BibitemOpen
  \bibfield  {author} {\bibinfo {author} {\bibfnamefont {J.~C.}\ \bibnamefont
  {Slater}},\ }\bibfield  {title} {\bibinfo {title} {A simplification of the
  hartree-fock method},\ }\href {https://doi.org/10.1103/PhysRev.81.385}
  {\bibfield  {journal} {\bibinfo  {journal} {Phys. Rev.}\ }\textbf {\bibinfo
  {volume} {81}},\ \bibinfo {pages} {385} (\bibinfo {year} {1951})}\BibitemShut
  {NoStop}%
\bibitem [{\citenamefont {Ros\'en}\ and\ \citenamefont
  {Lindgren}(1968)}]{RosenPR1968}%
  \BibitemOpen
  \bibfield  {author} {\bibinfo {author} {\bibfnamefont {A.}~\bibnamefont
  {Ros\'en}}\ and\ \bibinfo {author} {\bibfnamefont {I.}~\bibnamefont
  {Lindgren}},\ }\bibfield  {title} {\bibinfo {title} {Relativistic
  calculations of electron binding energies by a modified
  {H}artree-{F}ock-{S}later method},\ }\href
  {https://doi.org/10.1103/PhysRev.176.114} {\bibfield  {journal} {\bibinfo
  {journal} {Phys. Rev.}\ }\textbf {\bibinfo {volume} {176}},\ \bibinfo {pages}
  {114} (\bibinfo {year} {1968})}\BibitemShut {NoStop}%
\bibitem [{\citenamefont {Hahn}\ \emph {et~al.}(1956)\citenamefont {Hahn},
  \citenamefont {Ravenhall},\ and\ \citenamefont {Hofstadter}}]{HahnPR1956}%
  \BibitemOpen
  \bibfield  {author} {\bibinfo {author} {\bibfnamefont {B.}~\bibnamefont
  {Hahn}}, \bibinfo {author} {\bibfnamefont {D.~G.}\ \bibnamefont
  {Ravenhall}},\ and\ \bibinfo {author} {\bibfnamefont {R.}~\bibnamefont
  {Hofstadter}},\ }\bibfield  {title} {\bibinfo {title} {High-energy electron
  scattering and the charge distributions of selected nuclei},\ }\href
  {https://doi.org/10.1103/PhysRev.101.1131} {\bibfield  {journal} {\bibinfo
  {journal} {Phys. Rev.}\ }\textbf {\bibinfo {volume} {101}},\ \bibinfo {pages}
  {1131} (\bibinfo {year} {1956})}\BibitemShut {NoStop}%
\bibitem [{\citenamefont {Latter}(1955)}]{LatterPR1955}%
  \BibitemOpen
  \bibfield  {author} {\bibinfo {author} {\bibfnamefont {R.}~\bibnamefont
  {Latter}},\ }\bibfield  {title} {\bibinfo {title} {Atomic energy levels for
  the thomas-fermi and thomas-fermi-dirac potential},\ }\href
  {https://doi.org/10.1103/PhysRev.99.510} {\bibfield  {journal} {\bibinfo
  {journal} {Phys. Rev.}\ }\textbf {\bibinfo {volume} {99}},\ \bibinfo {pages}
  {510} (\bibinfo {year} {1955})}\BibitemShut {NoStop}%
\bibitem [{\citenamefont {Liberman}\ \emph {et~al.}(1971)\citenamefont
  {Liberman}, \citenamefont {Cromer},\ and\ \citenamefont
  {Waber}}]{LibermanCPC1971}%
  \BibitemOpen
  \bibfield  {author} {\bibinfo {author} {\bibfnamefont {D.}~\bibnamefont
  {Liberman}}, \bibinfo {author} {\bibfnamefont {D.}~\bibnamefont {Cromer}},\
  and\ \bibinfo {author} {\bibfnamefont {J.}~\bibnamefont {Waber}},\ }\bibfield
   {title} {\bibinfo {title} {Relativistic self-consistent field program for
  atoms and ions},\ }\href
  {https://doi.org/https://doi.org/10.1016/0010-4655(71)90020-8} {\bibfield
  {journal} {\bibinfo  {journal} {Computer Physics Communications}\ }\textbf
  {\bibinfo {volume} {2}},\ \bibinfo {pages} {107} (\bibinfo {year}
  {1971})}\BibitemShut {NoStop}%
\bibitem [{\citenamefont {Liberman}\ \emph {et~al.}(1965)\citenamefont
  {Liberman}, \citenamefont {Waber},\ and\ \citenamefont
  {Cromer}}]{LibermanPR1965}%
  \BibitemOpen
  \bibfield  {author} {\bibinfo {author} {\bibfnamefont {D.}~\bibnamefont
  {Liberman}}, \bibinfo {author} {\bibfnamefont {J.~T.}\ \bibnamefont
  {Waber}},\ and\ \bibinfo {author} {\bibfnamefont {D.~T.}\ \bibnamefont
  {Cromer}},\ }\bibfield  {title} {\bibinfo {title} {Self-consistent-field
  dirac-slater wave functions for atoms and ions. i. comparison with previous
  calculations},\ }\href {https://doi.org/10.1103/PhysRev.137.A27} {\bibfield
  {journal} {\bibinfo  {journal} {Phys. Rev.}\ }\textbf {\bibinfo {volume}
  {137}},\ \bibinfo {pages} {A27} (\bibinfo {year} {1965})}\BibitemShut
  {NoStop}%
\bibitem [{\citenamefont {Moliere}(1947)}]{MoliereZNA1947}%
  \BibitemOpen
  \bibfield  {author} {\bibinfo {author} {\bibfnamefont {G.}~\bibnamefont
  {Moliere}},\ }\bibfield  {title} {\bibinfo {title} {Theorie der streuung
  schneller geladener teilchen i. einzelstreuung am abgeschirmten
  coulomb-feld},\ }\href {https://doi.org/doi:10.1515/zna-1947-0302} {\bibfield
   {journal} {\bibinfo  {journal} {Zeitschrift für Naturforschung A}\ }\textbf
  {\bibinfo {volume} {2}},\ \bibinfo {pages} {133} (\bibinfo {year}
  {1947})}\BibitemShut {NoStop}%
\bibitem [{\citenamefont {Carlson}(1975)}]{CarlsonBook1975}%
  \BibitemOpen
  \bibfield  {author} {\bibinfo {author} {\bibfnamefont {T.~A.}\ \bibnamefont
  {Carlson}},\ }\href@noop {} {\emph {\bibinfo {title} {Photoelectron and
  {A}uger {S}pectroscopy}}}\ (\bibinfo  {publisher} {Plenum Press},\ \bibinfo
  {year} {1975})\BibitemShut {NoStop}%
\bibitem [{\citenamefont {Wang}\ \emph {et~al.}(2017)\citenamefont {Wang},
  \citenamefont {Audi}, \citenamefont {Kondev}, \citenamefont {Huang},
  \citenamefont {Naimi},\ and\ \citenamefont {Xu}}]{WangCPC2017}%
  \BibitemOpen
  \bibfield  {author} {\bibinfo {author} {\bibfnamefont {M.}~\bibnamefont
  {Wang}}, \bibinfo {author} {\bibfnamefont {G.}~\bibnamefont {Audi}}, \bibinfo
  {author} {\bibfnamefont {F.~G.}\ \bibnamefont {Kondev}}, \bibinfo {author}
  {\bibfnamefont {W.}~\bibnamefont {Huang}}, \bibinfo {author} {\bibfnamefont
  {S.}~\bibnamefont {Naimi}},\ and\ \bibinfo {author} {\bibfnamefont
  {X.}~\bibnamefont {Xu}},\ }\bibfield  {title} {\bibinfo {title} {The
  {AME}2016 atomic mass evaluation ({II}). tables, graphs and references},\
  }\href {https://doi.org/10.1088/1674-1137/41/3/030003} {\bibfield  {journal}
  {\bibinfo  {journal} {Chinese Physics C}\ }\textbf {\bibinfo {volume} {41}},\
  \bibinfo {pages} {030003} (\bibinfo {year} {2017})}\BibitemShut {NoStop}%
\bibitem [{\citenamefont {Wietfeldt}\ \emph {et~al.}(1995)\citenamefont
  {Wietfeldt}, \citenamefont {Norman}, \citenamefont {Chan}, \citenamefont
  {da~Cruz}, \citenamefont {Garc\'{\i}a}, \citenamefont {Haller}, \citenamefont
  {Hansen}, \citenamefont {Hindi}, \citenamefont {Larimer}, \citenamefont
  {Lesko}, \citenamefont {Luke}, \citenamefont {Stokstad}, \citenamefont
  {Sur},\ and\ \citenamefont {\ifmmode~\check{Z}\else
  \v{Z}\fi{}limen}}]{WietfeldtPRC1995}%
  \BibitemOpen
  \bibfield  {author} {\bibinfo {author} {\bibfnamefont {F.~E.}\ \bibnamefont
  {Wietfeldt}}, \bibinfo {author} {\bibfnamefont {E.~B.}\ \bibnamefont
  {Norman}}, \bibinfo {author} {\bibfnamefont {Y.~D.}\ \bibnamefont {Chan}},
  \bibinfo {author} {\bibfnamefont {M.~T.~F.}\ \bibnamefont {da~Cruz}},
  \bibinfo {author} {\bibfnamefont {A.}~\bibnamefont {Garc\'{\i}a}}, \bibinfo
  {author} {\bibfnamefont {E.~E.}\ \bibnamefont {Haller}}, \bibinfo {author}
  {\bibfnamefont {W.~L.}\ \bibnamefont {Hansen}}, \bibinfo {author}
  {\bibfnamefont {M.~M.}\ \bibnamefont {Hindi}}, \bibinfo {author}
  {\bibfnamefont {R.-M.}\ \bibnamefont {Larimer}}, \bibinfo {author}
  {\bibfnamefont {K.~T.}\ \bibnamefont {Lesko}}, \bibinfo {author}
  {\bibfnamefont {P.~N.}\ \bibnamefont {Luke}}, \bibinfo {author}
  {\bibfnamefont {R.~G.}\ \bibnamefont {Stokstad}}, \bibinfo {author}
  {\bibfnamefont {B.}~\bibnamefont {Sur}},\ and\ \bibinfo {author}
  {\bibfnamefont {I.}~\bibnamefont {\ifmmode~\check{Z}\else \v{Z}\fi{}limen}},\
  }\bibfield  {title} {\bibinfo {title} {Further studies on the evidence for a
  17-kev neutrino in a $^{14}\mathrm{doped}$ germanium detector},\ }\href
  {https://doi.org/10.1103/PhysRevC.52.1028} {\bibfield  {journal} {\bibinfo
  {journal} {Phys. Rev. C}\ }\textbf {\bibinfo {volume} {52}},\ \bibinfo
  {pages} {1028} (\bibinfo {year} {1995})}\BibitemShut {NoStop}%
\bibitem [{\citenamefont {Loidl}\ \emph {et~al.}(2020)\citenamefont {Loidl},
  \citenamefont {Beyer}, \citenamefont {Bockhorn} \emph
  {et~al.}}]{LoidlJLTP2020}%
  \BibitemOpen
  \bibfield  {author} {\bibinfo {author} {\bibfnamefont {M.}~\bibnamefont
  {Loidl}}, \bibinfo {author} {\bibfnamefont {J.}~\bibnamefont {Beyer}},
  \bibinfo {author} {\bibfnamefont {L.}~\bibnamefont {Bockhorn}}, \emph
  {et~al.},\ }\bibfield  {title} {\bibinfo {title} {Precision measurements of
  beta spectra using metallic magnetic calorimeters within the european
  metrology research project metrobeta},\ }\href
  {https://doi.org/10.1007/s10909-020-02398-2} {\bibfield  {journal} {\bibinfo
  {journal} {Journal of Low Temperature Physics}\ }\textbf {\bibinfo {volume}
  {199}},\ \bibinfo {pages} {451} (\bibinfo {year} {2020})}\BibitemShut
  {NoStop}%
\bibitem [{\citenamefont {Schopper}(1966)}]{Schopper-1966}%
  \BibitemOpen
  \bibfield  {author} {\bibinfo {author} {\bibfnamefont {H.~F.}\ \bibnamefont
  {Schopper}},\ }\href@noop {} {\emph {\bibinfo {title} {Weak Interactions and
  Nuclear $\beta$ Decay}}}\ (\bibinfo  {publisher} {North-Holland},\ \bibinfo
  {year} {1966})\BibitemShut {NoStop}%
\bibitem [{\citenamefont {Hayen}\ \emph {et~al.}(2020)\citenamefont {Hayen},
  \citenamefont {Simonucci},\ and\ \citenamefont {Taioli}}]{Hayen-arxiv2020}%
  \BibitemOpen
  \bibfield  {author} {\bibinfo {author} {\bibfnamefont {L.}~\bibnamefont
  {Hayen}}, \bibinfo {author} {\bibfnamefont {S.}~\bibnamefont {Simonucci}},\
  and\ \bibinfo {author} {\bibfnamefont {S.}~\bibnamefont {Taioli}},\ }\href
  {https://doi.org/10.48550/ARXIV.2009.08303} {\bibinfo {title} {Detailed
  $\beta$ spectrum calculations of $^{214}${P}b for new physics searches in
  liquid xenon}} (\bibinfo {year} {2020})\BibitemShut {NoStop}%
\bibitem [{\citenamefont {Mougeot}(2015)}]{MougeotPRC2015}%
  \BibitemOpen
  \bibfield  {author} {\bibinfo {author} {\bibfnamefont {X.}~\bibnamefont
  {Mougeot}},\ }\bibfield  {title} {\bibinfo {title} {Reliability of usual
  assumptions in the calculation of $\ensuremath{\beta}$ and $\ensuremath{\nu}$
  spectra},\ }\href {https://doi.org/10.1103/PhysRevC.91.055504} {\bibfield
  {journal} {\bibinfo  {journal} {Phys. Rev. C}\ }\textbf {\bibinfo {volume}
  {91}},\ \bibinfo {pages} {055504} (\bibinfo {year} {2015})}\BibitemShut
  {NoStop}%
\bibitem [{\citenamefont {Caprio}(2005)}]{SciDraw}%
  \BibitemOpen
  \bibfield  {author} {\bibinfo {author} {\bibfnamefont {M.}~\bibnamefont
  {Caprio}},\ }\bibfield  {title} {\bibinfo {title} {Levelscheme: A level
  scheme drawing and scientific figure preparation system for mathematica},\
  }\href {https://doi.org/https://doi.org/10.1016/j.cpc.2005.04.010} {\bibfield
   {journal} {\bibinfo  {journal} {Computer Physics Communications}\ }\textbf
  {\bibinfo {volume} {171}},\ \bibinfo {pages} {107 } (\bibinfo {year}
  {2005})}\BibitemShut {NoStop}%
\end{thebibliography}%

\end{document}